\documentclass[twocolumn,twocolappendix]{aastex631}
\usepackage{amsmath}
\usepackage{multirow}
\usepackage{enumitem}
\usepackage{color}
\usepackage{soul}
\usepackage{CJKfntef,CJK}
\usepackage{makecell}
\usepackage{graphicx}

\newcommand\kms{$\rm km\ s^{-1}$}
\newcommand\vlsr{$V_{\rm LSR}$}

\newcommand\hh{$\rm H_2$}
\newcommand\coa{$\rm ^{12}CO$}
\newcommand\cob{$\rm ^{13}CO$}
\newcommand\hcop{$\rm HCO^+$}
\newcommand\methanol{$\rm CH_3OH$}
\newcommand\emethanol{E-$\rm CH_3OH$}
\newcommand\amethanol{A-$\rm CH_3OH$}
\newcommand\phhco{$p$-$\rm H_2CO$}

\newcommand\hhco{$\rm H_2CO$}

\newcommand\sioonmethanol{$N({\rm SiO})/N({\rm CH_3OH})$}
\newcommand\soonmethanol{$N({\rm SO})/N({\rm CH_3OH})$}

\begin{document}
\begin{CJK*}{UTF8}{gbsn}
\title{An ACA map of a molecular cloud interacting with supernova remnant W28}

\author[0000-0002-9776-5610]{Tian-Yu Tu (涂天宇)}
\affiliation{School of Astronomy \& Space Science, Nanjing University, 163 Xianlin Avenue, Nanjing 210023, China}
\affiliation{Laboratoire d'Astrophysique de Bordeaux, Univ. Bordeaux, CNRS, B18N, allée Geoffroy Saint-Hilaire, 33615 Pessac, France}
\email{tianyu.tu1105@outlook.com}

\author[0000-0002-3599-6608]{Wenjin Yang (杨文锦)}
\affiliation{School of Astronomy \& Space Science, Nanjing University, 163 Xianlin Avenue, Nanjing 210023, China}

\author[0000-0002-4707-8409]{Siyi Feng (冯思轶)}
\affiliation{Department of Astronomy, Xiamen University, Zengcuo’an West Road, Xiamen, 361005}

\author[0000-0001-9676-2605]{Valentine Wakelam}
\affiliation{Laboratoire d'Astrophysique de Bordeaux, Univ. Bordeaux, CNRS, B18N, allée Geoffroy Saint-Hilaire, 33615 Pessac, France}

\author[0000-0002-4753-2798]{Yang Chen (陈阳)}
\affiliation{School of Astronomy \& Space Science, Nanjing University, 163 Xianlin Avenue, Nanjing 210023, China}
\affiliation{Key Laboratory of Modern Astronomy and Astrophysics, Nanjing University, Ministry of Education, Nanjing 210023, China}
\email{ygchen@nju.edu.cn}

\author[0000-0002-5683-822X]{Ping Zhou (周平)}
\affiliation{School of Astronomy \& Space Science, Nanjing University, 163 Xianlin Avenue, Nanjing 210023, China}
\affiliation{Key Laboratory of Modern Astronomy and Astrophysics, Nanjing University, Ministry of Education, Nanjing 210023, China}

\author[0000-0003-0853-1108]{Qian-Qian Zhang (张芊千)}
\affiliation{School of Astronomy \& Space Science, Nanjing University, 163 Xianlin Avenue, Nanjing 210023, China}

\begin{abstract}
Supernova remnants (SNRs) strongly influence the physical and chemical properties of the molecular clouds (MCs) with which they interact. 
We carried out a high-resolution observation toward W28F, a chemically rich MC interacting with SNR W28, with the Atacama Compact Array (ACA) in Band 7. 
Significant emission ($>10\sigma$) of CO, \methanol, \phhco, SiO and SO is detected. 
We reveal the clumpy structures of the shocked MC, with different spatial distributions between \methanol\ and SiO. 
We select six molecular clumps to conduct spectral decomposition and non-local-thermodynamic-equilibrium analysis with the \methanol\ and \phhco\ lines. 
The best-fit results show a \hh\ density of $n_{\rm H_2}\sim1$--$3\times10^5\rm \ cm^{-3}$ and a gas temperature of $T_{\rm gas}\sim50$--170 K in most of the fitted components. 
The \hh\ density and gas temperature show a clear anti-correlation across different regions, with the thermal pressure consistent with that of the adjacent X-ray-emitting hot plasma. 
This is consistent with the picture that the SNR shocks propagate into multi-phase gas, with a pressure balance existing between different phases.
We propose that the high abundance ratio between \emethanol and \amethanol\ ($>0.9$) suggests extra gas-phase processes to enhance this ratio, such as proton exchange with $\rm H_3^+$ and \hcop. 
The chemical segregation between \methanol\ and SiO, in both the spatial and spectral regime, can be explained by the fact that \methanol\ traces slow shocks while SiO traces fast shocks. 

\end{abstract}

\keywords{Molecular clouds (1072) --- Supernova remnants (1667) ---  Astrochemistry (75) --- Shocks (2086) --- Abundance ratios (11)}

\section{Introduction} \label{sec:intro}

Supernova remnants (SNRs) exert strong influence on the physical and chemical properties of the molecular clouds (MCs) with which they interact, regulating star formation and galaxy evolution \citep[e.g.,][]{Kim_Momentum_2015,Pillepich_Simulating_2018}. 
So far, in our Galaxy where the SNR-MC interaction can be spatially resolved, tens of SNRs have been found to be interacting with adjacent MCs \citep[e.g.,][]{Jiang_Cavity_2010,Zhou_Systematic_2023}. 
The physical parameters of these MCs can be significantly altered by the heating and compression effect of shocks \citep{Flower_Theoretical_1985,Draine_Theory_1993}. 
The X-ray emission from the SNRs and the cosmic-rays (CRs) accelerated by the SNR shocks are also important heating source of MCs \citep{Meijerink_Irradiated_2006}. 
Molecular line observations which are able to estimate the physical parameters of the shocked MCs have confirmed the physical feedback of SNRs \citep{vanDishoeck_Submillimeter_1993,Reach_Excitation_1999,Anderl_APEX_2014,Hogge_Interaction_2019,DellOva_Interstellar_2020}. 

\par

SNRs can also change the chemical properties, including molecular abundances and abundance ratios, of the associated MCs. 
So far, a series of chemical effects related to SNR-MC interaction have been found. They include (1) formation of SiO due to shock sputtering of dust grains \citep{vanDishoeck_Submillimeter_1993,Nicholas_7_2012,Cosentino_Interstellar_2019,Cosentino_Negative_2022}, (2) release of dust mantle species to the gas phase (e.g., $\rm NH_3$ \citep{Gusdorf_Probing_2012} and \hhco\ \citep{Mazumdar_Submillimeter_2022a}), (3) enriched $\rm H_3^+$ molecules and reduced deuteration of ionized species such as \hcop\ due to CR ionization \citep{Indriolo_Investigating_2010,Ceccarelli_Supernova-enhanced_2011,Vaupre_Cosmic_2014,Indriolo_Absorption-line_2023}, (4) transformation from CO to atomic C induced by $\rm He^+$ ionized by CRs \citep{Yamagishi_Cosmic-ray-driven_2023}, (5) enhanced $N({\rm HOC^+})/N({\rm HCO^+})$ and $N({\rm HCS^+})/N({\rm CS})$ due to CR-induced chemistry \citep{Tu_Yebes_2024}, (6) enhanced $N({\rm HCO^+})/N(\rm CO)$ due to both shock and CRs \citep{Zhou_Unusually_2022b,Tu_Shock_2024}, etc. 
In dense star-forming regions, molecular species whose abundances are sensitive to shock interaction are regarded as shock tracers and used to study the properties of shock wave. 
These species, including SiO \citep{Martin-Pintado_SiO_1992}, \methanol\ \citep{Bachiller_Methanol_1995}, HNCO \citep{Rodriguez-Fernandez_HNCO_2010}, etc., are mainly formed in dust grains and can be released to the gas phase by shocks via sputtering and grain-grain collision. However, they are seldom used to study the shocked MCs by SNRs because of the limited sensitivity of previous observations. 
They have only been detected in a limited number of MCs interacting with SNRs --- IC443 \citep{vanDishoeck_Submillimeter_1993}, W28 \citep{Nicholas_7_2012,Mazumdar_Submillimeter_2022a}, and W44 \citep{Cosentino_Widespread_2018,Cosentino_Interstellar_2019}. 
By taking the advantage of these shock tracers, we can study the properties of the shocked MCs without the confusion of unrelated velocity components in the line of sight.

\par

SNR W28 is one of the best-studied SNRs interacting with MCs. 
It is believed to be interacting with the dense MCs located toward its northeastern edge, which is evidenced by broadened (with $\rm FWHMs \gtrsim 10$ \kms) molecular lines \citep{Wootten_dense_1981,Arikawa_Shocked_1999,Reach_Shocked_2005,Nicholas_12_2011,Nicholas_7_2012,Gusdorf_Probing_2012,Maxted_Ammonia_2016,Mazumdar_Submillimeter_2022a,Tu_Shock_2024}, 1720 MHz OH masers \citep{Claussen_Polarization_1997,Hoffman_OH_2005}, class I \methanol\ masers \citep{Pihlstrom_Detection_2014}, and infrared emission from \hh\ and $\rm H_2O$ \citep{Reach_Shocked_2005,Yuan_Spitzer_2011,Neufeld_Water_2014}. 
Recent studies have revealed the chemical effects of W28 on its northeastern MCs. 
By investigating the ortho-to-para ratios of $\rm NH_3$ and \hhco, \citet{Maxted_Ammonia_2016} and \citet{Mazumdar_Submillimeter_2022a} found that these two molecular species are released to the gas phase because of shocks with velocities of $\sim 15$--25 \kms\ \citep{Gusdorf_Probing_2012,Tu_Shock_2024} that heat the gas to $\sim 40$--100 K. 
\citet{Tu_Shock_2024} found enhanced $N({\rm HCO^+})/N(\rm CO)$ abundance ratio by an order of magnitude in shocked MCs and attributed it to the chemical effect of both shocks and CRs. 

\begin{figure}[tbp]
\centering
\includegraphics[width=0.47\textwidth]{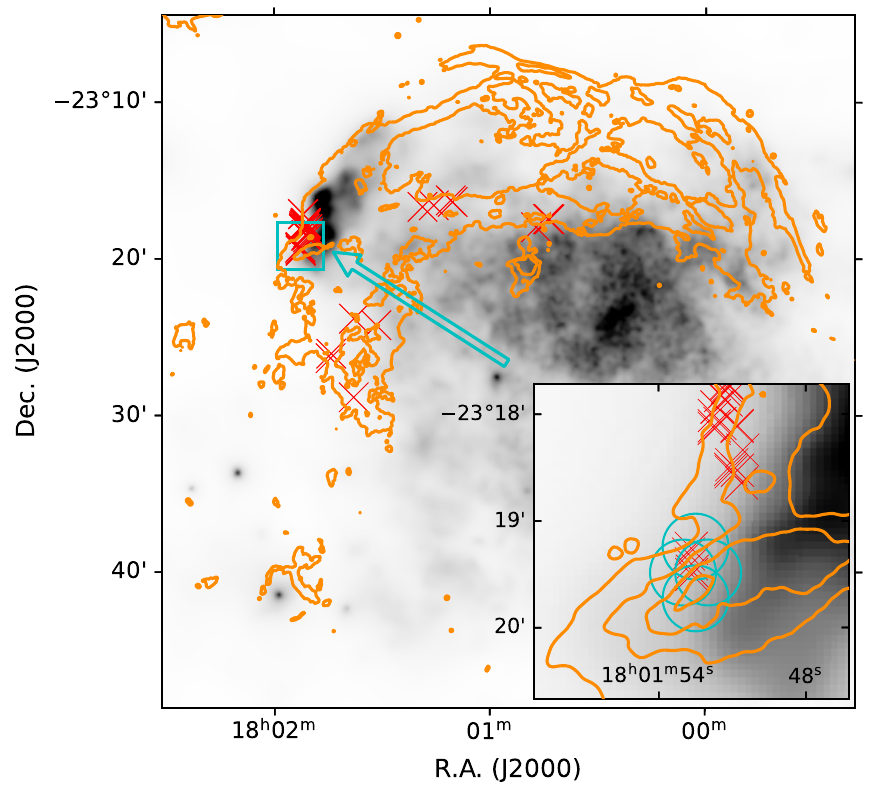}
\caption{X-ray image of SNR W28 observed by the Follow-up X-ray Telescope (FXT) onboard the Einstein Probe (EP) in 0.4--2.3 keV band \citep{Chi_complete_2026}, overlaid with contours of MeerKAT 1.3 GHz radio continuum emission  \citep{Goedhart_SARAO_2024} in levels of 3 and 10 mJy $\rm beam^{-1}$. 
The red crosses denote the 1720 MHz OH masers detected by \citet{Claussen_Polarization_1997}. 
A zoom-in view around W28F is shown in the inset figure, with the contour levels in 4, 7, 10 and 13 mJy $\rm beam^{-1}$. 
The cyan circles mark the primary beams of the four pointings that constitute the mosaic centered at the MC W28F.
\label{fig1}
}
\end{figure}

W28F, a region toward the northeastern edge of W28, was named by \citet{Claussen_Polarization_1997} who detected several 1720 MHz OH masers. 
Two 36 GHz Class I \methanol\ masers were also found in W28F \citep{Pihlstrom_Detection_2014}. 
We hereafter refer W28F to the MC associated with these masers \citep{Mazumdar_Submillimeter_2022a}. 
In Figure \ref{fig1} we show its position. 
It is also located right outside an ear-like X-ray structure \citep[e.g.,][]{Rho_ROSAT_2002,Zhou_XMM-Newton_2014,Chi_complete_2026} and some association between them has been suggested \citep{Nicholas_7_2012}. 
Recent sub-millimeter line survey carried out by \citet{Mazumdar_Submillimeter_2022a} toward W28F using the Atacama Pathfinder EXperiment (APEX) telescope revealed its rich chemical inventory, including \hhco, \methanol, SiO, SO, $\rm N_2H^+$, etc. 
However, all of the previous observations pointed out that W28F is too clumpy ($\lesssim 30^{\prime\prime}$ in diameter) to be resolved by single-dish telescopes \citep{Nicholas_7_2012,Mazumdar_Submillimeter_2022a,Tu_Shock_2024}. 

\par

In this paper, we report our new observation toward W28F with the Atacama Compact Array (ACA) delivering spatially resolved information that allows for an in-depth chemical analysis in W28F. 
The paper is organized as follows. 
In Section \ref{sec:obs}, we briefly introduce our ACA observation. 
We present the results of the observation, including the intensity maps of the detected transition lines, in Section \ref{sec:res}. 
Then, we estimate the physical parameters, and analyze the chemical property from \methanol, SiO, \hhco, and SO in Section \ref{sec:disc}. 
Our main findings are summarized in Section \ref{sec:con}.

\section{Observation} \label{sec:obs}

We carried out an observation toward SNR W28 with ACA and the Total Power (TP) array of the Atacama Large Millimeter/sub-millimeter Array (ALMA) in band 7 during Cycle 11 (PI: Tian-Yu Tu, \#2024.1.00194.S). 
A small mosaic of 4 pointings was made to cover the W28F region, centered at $\rm R.A.=18^h01^m52^s.5$, $\rm Dec.=-23^\circ19^\prime29^{\prime\prime}.0$ (see Figure \ref{fig1}), corresponding to the \amethanol\ ($6_0$--$5_0$)\footnote{Throughout this paper, we follow the label listed in the Leiden Atomic and Molecular Database (LAMDA) database \citep{vanderTak_Leiden_2020} to denote the transitions.} emission peak where \citet{Mazumdar_Submillimeter_2022a} carried out the line survey. 
The observation consists of two tunings, one mainly targeting the $J=3$--2 transition of \coa\ at 345.79599~GHz and \cob\ at 330.58796~GHz (with a beam size of $\sim 4^{\prime\prime}.3\times 3^{\prime\prime}.2$, corresponding to $\rm 0.040\ pc \times0.029\ pc$  at the distance of W28 \citep[$\sim 1.9$ kpc,][]{Velazquez_Investigation_2002}), and the other mainly targeting the \methanol, \phhco, and SiO lines around 290~GHz (with a beam of $\sim 5^{\prime\prime}.3\times 3^{\prime\prime}.1$, corresponding to $\rm 0.049\ pc \times0.029\ pc$). 
We also placed three continuum spectral windows with widths of 1875~MHz around 304.0, 331.0, and 344.0~GHz. 
The first tuning was observed in December 2024, while the second was observed in April--May 2025.
All the detected molecular transitions and their velocity channel widths are listed in Table \ref{tab:lines}. 
The data were calibrated and imaged with Briggs robust weighting of 0.5 by the standard reduction pipeline with the Common Astronomy Software Applications package \citep[CASA, version: 6.6.1,][]{CASATeam_CASA_2022}. 
To recover the large-scale emission, we combined the ACA and TP data using the \texttt{Feather} task in CASA with the parameter \texttt{sdfactor} set to be 1, a commonly used value \citep[e.g.,][]{Plunkett_Data_2023a}. 
The resulting synthesized beams and typical noise values of the detected transitions are summarized in Table \ref{tab:lines}. 
We did not detect any continuum emission in any of the three continuum spectral windows. 

\par

To facilitate our analysis and data visualization, we also retrieved 
the radio continuum map of SNR W28 from the SARAO MeerKAT 1.3 GHz Galactic Plane Survey with an angular resolution of $8^{\prime\prime}$ \citep{Goedhart_SARAO_2024}. 
All the data were further analyzed with \texttt{Python} packages \texttt{Astropy} and \texttt{Spectral-cube}, and visualized by \texttt{Matplotlib}. 

\begin{deluxetable*}{cccccccc}[ht]
\tablecaption{A summary of the basic information of the detected lines. 
\label{tab:lines}}
\tablehead{ 
\colhead{Species} & 
\colhead{Transition} & 
\colhead{\makecell{Frequency \\ (GHz)}} & 
\colhead{\makecell{$E_{\rm up}$ \\ (K)}} & 
\colhead{\makecell{$n_{\rm cr}$$\rm ^a$ \\ $\rm (cm^{-3})$}} & 
\colhead{\makecell{Beam, P.A. \\ ($^{\prime\prime}\times^{\prime\prime}$, $^{\circ}$)}} &
\colhead{\makecell{Channel Width \\ (\kms)} } &
\colhead{\makecell{Noise$\rm ^b$ \\ ($\rm mK\ chan^{-1}$)} }
}
\startdata 
\coa\ & 3--2 & 345.795990 & 33.2 & $8.2\times10^3$ & $4.30\times 3.17,\ 74.4$ & 0.21 & 84 \\ 
\cob\ & 3--2 & 330.687965 & 31.7 & $7.0\times10^3$ & $4.62\times 3.17,\ 70.8$ & 0.22 & 86 \\ 
\emethanol\ & $6_0$--$5_0$ & 289.939377 & 61.8 & $4.4\times10^5$ & $5.33\times 3.12,\ -85.9$ & 0.50 & 12 \\ 
& $6_{-1}$--$5_{-1}$ & 290.069747 & 54.3 & $1.4\times10^5$  & $5.33\times 3.12,\ -85.9$ & 0.50 & 12 \\ 
& $6_1$--$5_1$ & 290.248685 & 69.8 & $5.3\times10^5$ & $5.33\times 3.12,\ -85.9$ & 0.50 & 12 \\ 
& $6_2$--$5_2$ & 290.307738 & 71.0 & $2.7\times10^5$ & $5.33\times 3.12,\ -85.9$ & 0.50 & 12 \\ 
& $3_0$--$2_{-1}$ & 302.369773 & 27.1 & $1.0\times10^5$ & $5.13\times 2.95,\ -86.8$ & 0.48 & 14 \\ 
\amethanol\ & $6_0$--$5_0$ & 290.110637 & 48.7 & $1.3\times10^5$ & $5.33\times 3.12,\ -85.9$ & 0.50 & 12 \\ 
& $6_{-3}$--$5_{-3}$ $\rm ^c$ & 290.190549 & 98.5 & $1.2\times10^6$ & $5.33\times 3.12,\ -85.9$ & 0.50 & 12 \\ 
\phhco\ & $4_{0,4}$--$3_{0,3}$ & 290.623405 & 34.9 & $9.3\times10^5$ & $5.34\times 3.11,\ -85.9$ & 0.50 & 9 \\ 
& $4_{2,3}$--$3_{2,2}$ & 291.237766 & 82.1 & $6.7\times10^5$  & $5.33\times 3.12,\ -86.0$ & 0.50 & 10 \\ 
& $4_{2,2}$--$3_{2,1}$ & 291.948067 & 82.1 & $6.8\times10^5$  & $5.35\times 3.09,\ -86.2$ & 0.50 & 10 \\ 
SO & $7_8$--$6_7$ & 304.077844 & 62.1 & $1.0\times10^6$ & $5.08\times 2.94,\ -85.3$ & 0.48 & 12 \\ 
& $8_8$--$7_7$ & 344.310612 & 87.5 & $1.5\times10^6$  & $4.32\times 3.20,\ 72.0$ & 0.21 & 76 \\ 
$\rm ^{34}SO$ & $7_6$--$6_5$ & 290.562894 & 63.8 & $9.4\times10^5$ $\rm ^d$ & $5.34\times 3.11,\ -85.9$ & 0.50 & 9 \\ 
SiO & 7--6 & 303.926960 & 58.3 & $1.7\times10^6$ & $5.10\times 2.93,\ -85.6$ & 0.48 & 14 \\ 
OCS & 24--23 & 291.839654 & 175.1 & $1.4\times10^5$ & $5.34\times 3.09,\ -86.6$ & 0.50 & 12 \\ \hline
\enddata
\tablecomments{
$^{\rm a}$ The critical densities of the selected transition were calculated using the method described by \citet{Shirley_Critical_2015}, assuming optically thin emission and a gas kinetic temperature of 80 K.
The molecular data are retrieved from the Excitation of Molecules and Atoms for Astrophysics (EMAA, \url{https://emaa.osug.fr/}) database, except for SO from the LAMDA database \citep{vanderTak_Leiden_2020}.
$^{\rm b}$ Noise estimated at the map center of the selected transition converted from the unit of $\rm mJy\ beam^{-1}\ chan^{-1}$ at the referred beam size and channel width. 
The averaged noise level throughout the map can be estimated as $\sim2$--2.5 times the noise at the map center. 
$^{\rm c}$ The \amethanol\ $6_{-3}$--$5_{-3}$ is overlapped with its $6_{3}$--$5_{3}$ line with a rest frequency of 290189.515 MHz and similar properties. Since the target MCs exhibit broadened line profiles, we could not distinguish between these two lines. 
$^{\rm d}$ The collisional rate coefficient of $\rm ^{34}SO $ is currently not available. Here we use the value of SO as an approximation. 
}
\end{deluxetable*}

\section{Results} \label{sec:res}

\subsection{Integrated intensity maps and spatial distribution of the detected species} \label{sec:integrated_intensity}

\begin{figure*}[htbp]
\centering
\includegraphics[width=0.99\textwidth]{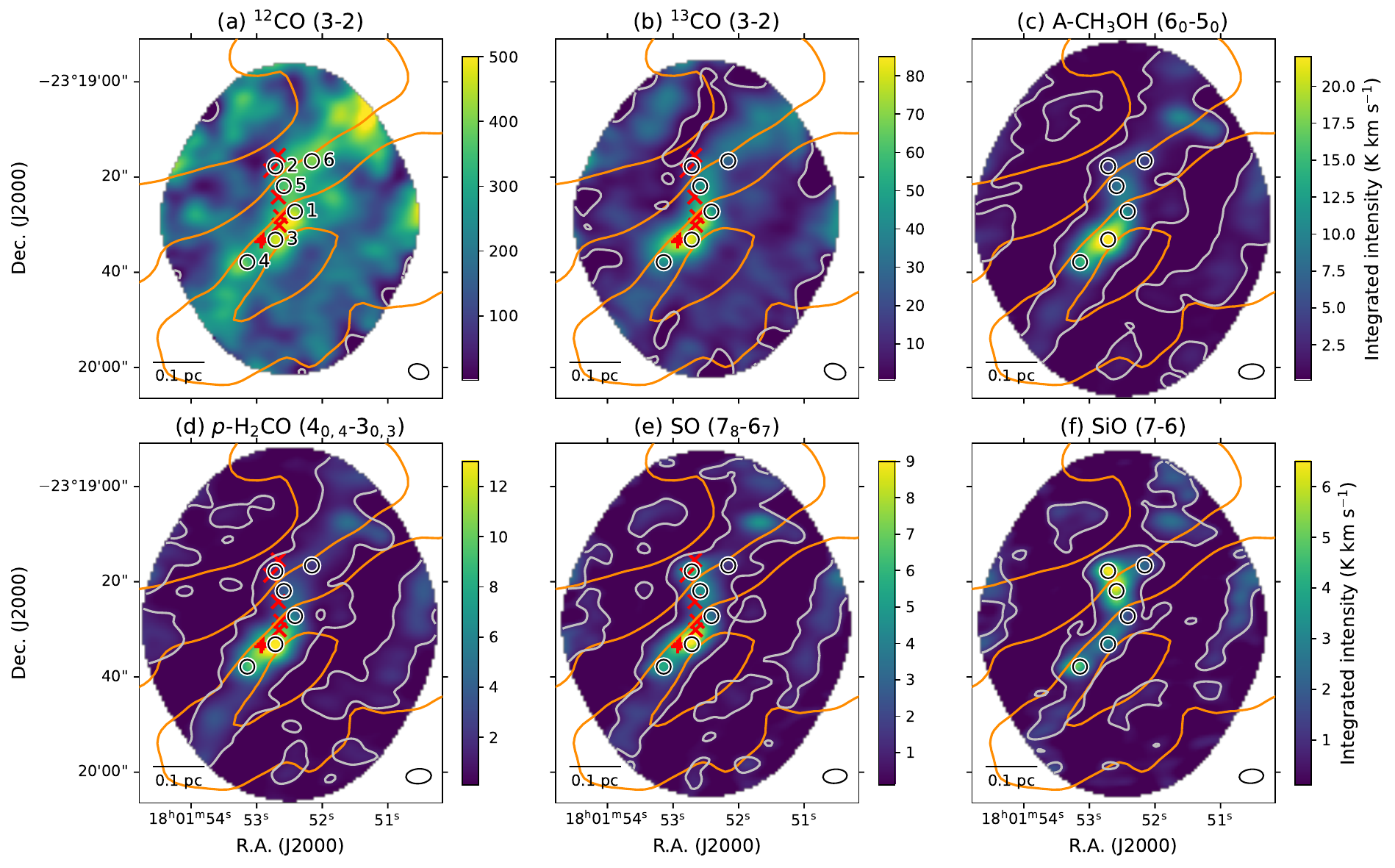}
\caption{Integrated intensity maps of \coa\ (3--2) (a), \cob\ (3--2) (b), \amethanol\ ($6_0$--$5_0$) (c), \phhco\ ($4_{0,4}$--$3_{0,3}$) (d), SO ($7_8$--$6_7$) (e), and SiO (7--6) (f) in $-12$ to $+20$ \kms, overlaid with orange contours of MeerKAT 1.3 GHz radio continuum emission in levels of 4, 7, 10 and 13 mJy $\rm beam^{-1}$, and grey contours showing the $5\sigma$ detection limit. 
The black circles are the regions where we extract the spectra. 
How we choose these regions is explained in Section \ref{sec:velocity_channel_map}. 
The red crosses and plus signs in (a), (b), (d) and (f) show the 1720 MHz masers detected by \citet{Claussen_Polarization_1997} and the two (overlapping) 36 GHz \methanol\ masers detected by \citet{Pihlstrom_Detection_2014}, respectively. 
The beam size of each transition is shown as an open black ellipse at the lower right of each sub-figure. 
\label{fig:integrated_intensity_1}
}
\end{figure*}

\begin{figure*}[htbp]
\centering
\includegraphics[width=0.97\textwidth]{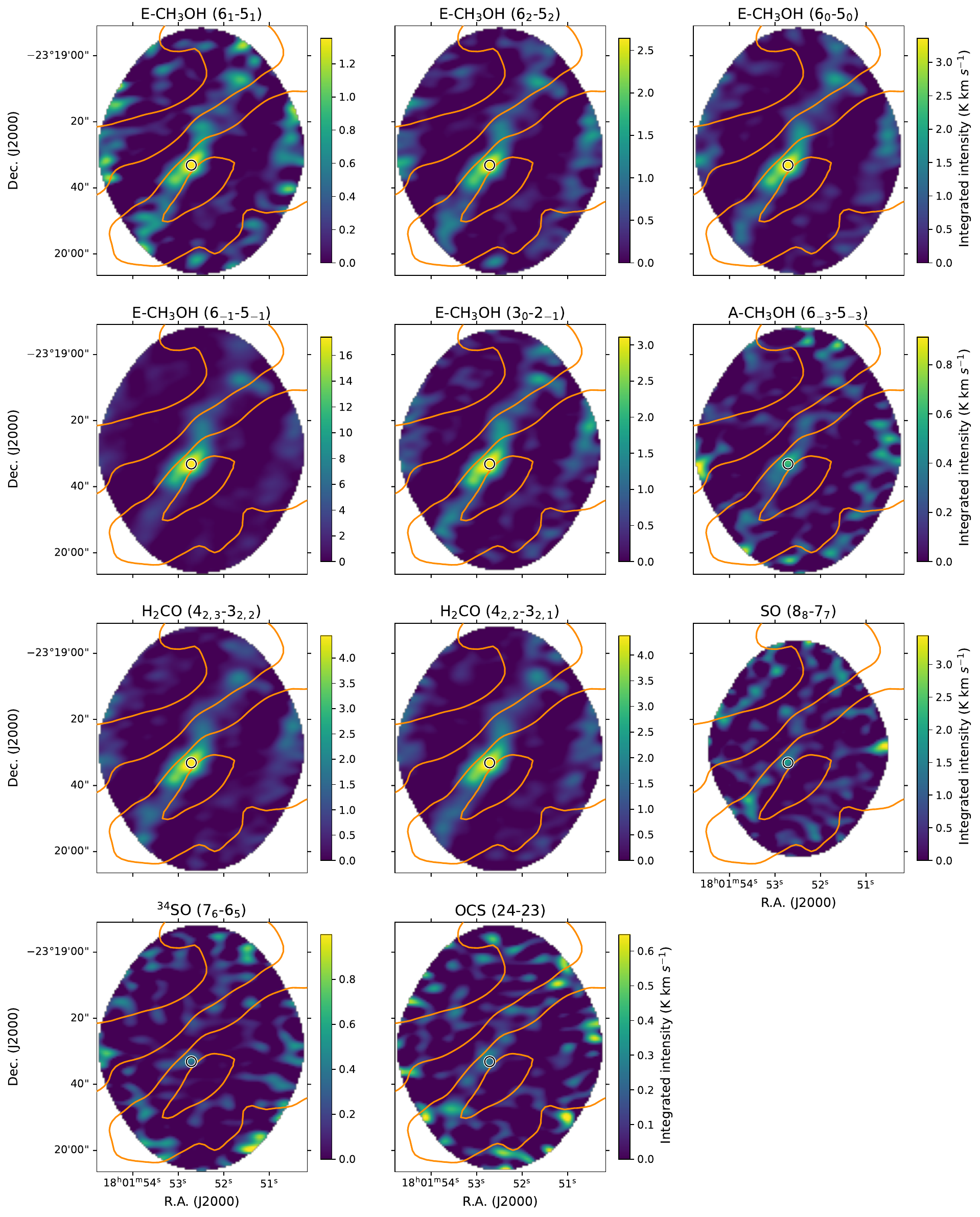}
\caption{Integrated intensity maps in $-12$ to $+20$ \kms of all detected transitions other than those shown in Figure \ref{fig:integrated_intensity_1}. 
The orange contours are the 1.3 GHz radio continuum, while the black circle shows the position of region 3. 
\label{fig:integrated_intensity_2}
}
\end{figure*}

In Figure \ref{fig:integrated_intensity_1} we show the integrated intensity maps of \coa\ (3--2), \cob\ (3--2), \amethanol\ ($6_0$--$5_0$), \phhco\ ($4_{0,4}$--$3_{0,3}$), SO ($7_8$--$6_7$), and SiO (7--6) over the velocity range $-12$--$+20$ \kms. This velocity range covers all of the emission associated with W28.
All of the six selected transitions exhibit extended emission with bright knots with angular sizes close to the beam size. 
Generally, these transitions show similar spatial distribution close to the radio continuum of the SNR. 
The line emission of molecular species are all located right outside the northeastern boundary of a radio knot, while the 1720 MHz OH masers, in turn, outlines the northeastern boundary of the molecular emission. 
Six regions with radii of $1^{\prime\prime}.5$ are marked in Figure \ref{fig:integrated_intensity_1}, which we focus on for subsequent analysis. 
Their selections are explained in \ref{sec:velocity_channel_map} and their coordinates are listed in Table \ref{tab:coord_regs}. 

{\color{blue}
\begin{deluxetable}{ccc}[ht]
\tablecaption{Center coordinates of the six regions where we extract the spectra. 
\label{tab:coord_regs}}
\tablehead{ 
\colhead{Region} & 
\colhead{R.A. (J2000)} & 
\colhead{Dec. (J2000)} 
}
\startdata 
1 & 18:01:52.41 & $-$23:19:27.2 \\
2 & 18:01:52.71 & $-$23:19:17.8 \\
3 & 18:01:52.71 & $-$23:19:33.1 \\
4 & 18:01:53.14 & $-$23:19:37.8 \\
5 & 18:01:52.59 & $-$23:19:21.9 \\
6 & 18:01:52.16 & $-$23:19:16.6 \\
\hline
\enddata
\end{deluxetable}
}

\par

The \coa\ (3--2) line reveals the most extended emission, with its peak located between regions 1 and 3. 
On the other hand, the \cob\ (3--2) line traces more compact emission peaking at region 3, coinciding with the peak position of the \amethanol, \phhco, and SO emission. 
These three molecular species, together with SiO, primarily trace the clumpy structures, but their spatial distributions are not identical. 
The \amethanol\ and \phhco\ lines are spatially coincident with each other, while the SiO (7--6) emission exhibits three prominent peaks at regions 2, 4, and 5. 
Although these three regions also show emission from \amethanol\ and \phhco, they are less prominent than the SiO emission. 
The SO emission is partially similar to both \amethanol\ and SiO in spatial distribution --- it peaks at region 3, but also has clumpy unresolved structures at regions 2 and 5. 

\par

The integrated intensity maps of the other detected species are shown in Figure \ref{fig:integrated_intensity_2}. 
The \methanol\ and \phhco\ emissions generally show similar spatial distribution as the \amethanol\ ($6_0$--$5_0$) and the \phhco\ ($4_{0,4}$--$3_{0,3}$) lines (see panels c and d of Figure \ref{fig:channel_maps}), peaking at region 3, albeit with lower brightness. 
The SO ($8_8$--$7_7$), $\rm ^{34}SO$ ($7_6$--$6_5$), and OCS (24--23) lines are only marginally detected at regions 3 ($\sim 3$, 7, and 8$\sigma$, respectively). 

\subsection{Velocity channel maps of \amethanol\ and SiO} \label{sec:velocity_channel_map}

\begin{figure*}[htbp]
\centering
\includegraphics[width=0.90\textwidth]{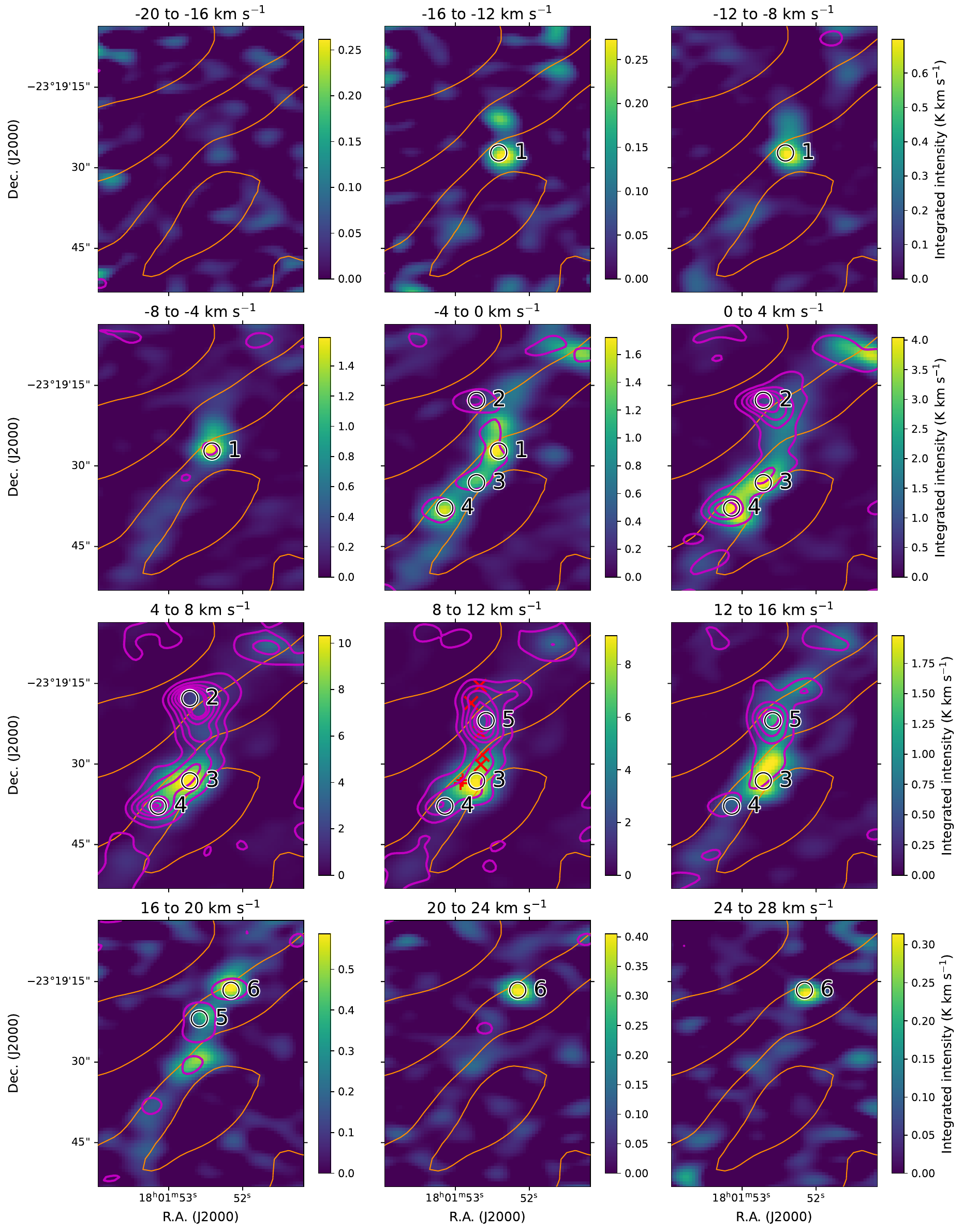}
\caption{Integrated intensity maps of \amethanol\ ($6_0$--$5_0$) line in each 4 \kms\ interval in the velocity range $-20$ to $+28$ \kms, overlaid with orange contours of 1.3 GHz radio continuum. 
The magenta contours are the integrated intensity of the SiO (7--6) line in the same velocity interval in levels of 5, 15, 25, 35, 45, and 50 times the typical noise level ($\sim 0.039\rm \ K\ km\ s^{-1}$). 
The black circles mark the regions the same as those shown in Figure \ref{fig:integrated_intensity_1}. 
The red crosses and plus signs in map $+8$ to $+12$ \kms\ stand for the 1720 OH masers and the 36 GHz \methanol\ maser, respectively. 
\label{fig:channel_maps}
}
\end{figure*}

To further investigate the different spatial distributions of \amethanol\ ($6_0$--$5_0$) and SiO (7--6) at different velocity intervals, Figure \ref{fig:channel_maps} shows the integrated intensity maps in steps of 4~\kms\ over the velocity range from $-20$ to $+28$~\kms\ to cover the entire velocity interval that contains emission.
These maps reveal that the clumpy structures peak at different velocity intervals.  
The regions 1--6 are chosen from the molecular line emissions in this figure sorted by ascending \vlsr. 
Emission in Region 1 stands out in negative \vlsr\ range ($-16$ to 0 \kms) at the center of the map, while emission in region 6 is prominent in the most positive \vlsr\ range ($+16$ to $+28$ \kms) located at the northwest of the map. 
Regions 2 and 5 are selected based on the clumpy SiO emission peaking at 0 to $+8$ \kms\ and $+8$ to $+16$ \kms, respectively, which are the most prominent components in the integrated intensity map of SiO (panel f of Figure \ref{fig:integrated_intensity_1}). 
In region 3, the peak of \amethanol\ emission at $+4$ to $+8$ \kms\ corresponds to the brightest emission of all \methanol, \phhco, and SO lines shown in Figures \ref{fig:integrated_intensity_1} and \ref{fig:integrated_intensity_2}. 
Region 4, located southeast of region 3, exhibits both significant \amethanol\ and SiO emission in $-4$ to $+16$ \kms. 

\par

We note that in Figure \ref{fig:channel_maps}, even though we focus on a small field of view around the selected clumps, the peak positions of the emission vary across different velocity intervals. 
For example, the \amethanol\ emission peak is spatially coincident with region 1 in $-8$ to 0 \kms, but is slightly southwest to region 1 in $-16$ to $-8$ \kms. 
This is suggestive of the complicated kinematics of the shocked molecular clumps.

\par

The segregation between \methanol\ and SiO is more prominent in these channel maps.
The most significant segregation appears at region 2, where the SiO emission peaks while the \amethanol\ emission is faint and the majority of the \amethanol\ emission arises from the western side of the SiO clump. 
Region 5 is also bright in SiO line but is closer to the main \amethanol\ emission. 
We also note that although clumpy \amethanol\ and SiO are both apparent at region 4, their spatial distributions show slight difference: the SiO emission is located to the east of the \amethanol\ emission. 

\par

Five 1720 MHz OH masers detected by \citet{Claussen_Polarization_1997} are covered in our map. 
The \vlsr\ of these masers range from +8.58 to +11.66 \kms.
Therefore, our channel map in $+8$ to $+12$ \kms\ rightly cover the velocity range of these OH masers. 
Although both \amethanol\ and SiO show emission peaks (regions 2 and 5, respectively) in this velocity range, the OH masers exhibit better spatial coincidence with the SiO emission than the \amethanol\ emission in $+8$ to $+12$ \kms\ of Figure \ref{fig:channel_maps}. 
In addition, the brightest OH maser spot appears to be located closest to the peak of the integrated SiO emission (region 5), and the second strongest OH maser appears to be the second closest to the peak.
On the contrary, the two 36 GHz \methanol\ masers detected by \citet{Pihlstrom_Detection_2014} at +8.2 \kms, are found to be offset from the \methanol\ peak (region 3).

\subsection{Spectra of \coa, \cob, \amethanol\ and SiO toward the six selected regions} \label{sec:result_spec}

\begin{figure*}[!t]
\centering
\includegraphics[width=0.95\textwidth]{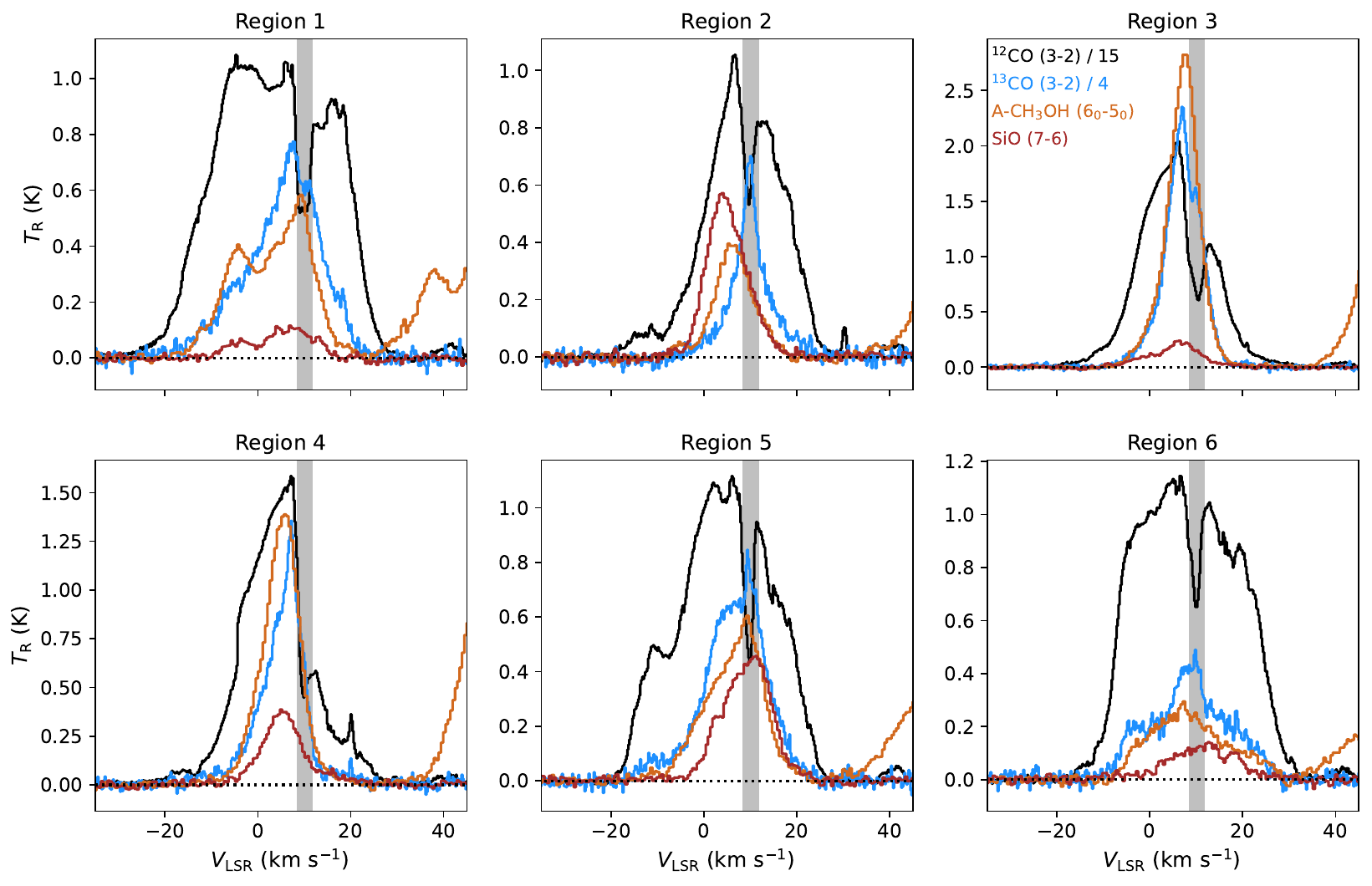}
\caption{Spectra of the \coa\ (3--2, in black, scaled by a factor of 1/15) and \cob\ (3--2, in blue, scaled by a factor of 1/4), \amethanol\ ($6_0$--$5_0$, in orange), and SiO (7--6, in brown) lines in the six selected regions. 
The gray shaded regions show the range of the \vlsr\ of the 1720 MHz OH masers. 
The emission in the \amethanol\ spectra at $V_{\rm LSR}\gtrsim25\rm \ km\ s^{-1}$ is from the \emethanol\ ($6_{-1}$--$5_{-1}$) line. 
\label{fig:COspec}
}
\end{figure*}

Figure \ref{fig:COspec} shows the spectra of the \coa\ (3--2), \cob\ (3--2), \amethanol\ ($6_0$--$5_0$), and SiO (7--6) lines extracted from the six selected regions. 
The \coa\ (3--2) spectra exhibit complex and broadened line profiles due to shock interaction, with contamination from unrelated components along the lines of sight.
Self-absorption of \coa\ (3--2) can be recognized in all the six regions, which has also been reported in \citet{Gusdorf_Probing_2012} and \citet{Mazumdar_Submillimeter_2022a}. 
The \vlsr\ of the self-absorption dips are consistent with the \vlsr\ range of the 1720 MHz OH masers \citep{Claussen_Polarization_1997}, and also with those of the emission peaks of the \cob\ (3--2) lines except for regions 4. 
This consistency suggests that the \coa\ self-absorption and the \cob\ peak likely originate from the preshock gas toward W28F. 
The line profiles of the \cob, \amethanol, and SiO lines are less affected by the unrelated components than \coa. 
Generally, the profiles of the \cob\ and \amethanol\ spectra are similar in the red wings, but show large difference in the blue wings, except for region 3. 
The line profiles of \cob\ with broad wings show that it traces not only the preshock but also the shocked gas. 
The \amethanol\ and SiO lines are red-shifted from the \coa\ absorption dip except in regions 5 and 6. 
Significant difference between the line profiles of \cob, \amethanol\ and SiO is found in region 2, i.e. the SiO emission peak, with \amethanol\ significantly blue-shifted compared with the \cob\ peak and SiO even more blue-shifted. 
The detected lines exhibit complex velocity structures that vary across different regions toward W28F, indicating intricate gas kinematics induced by SNR shock waves.

\section{Discussion} \label{sec:disc}

\subsection{(Multi-)Gaussian line decomposition of the spectra in the six selected regions}
\label{sec:decomposition}

\begin{figure*}[!t]
\centering
\includegraphics[width=0.95\textwidth]{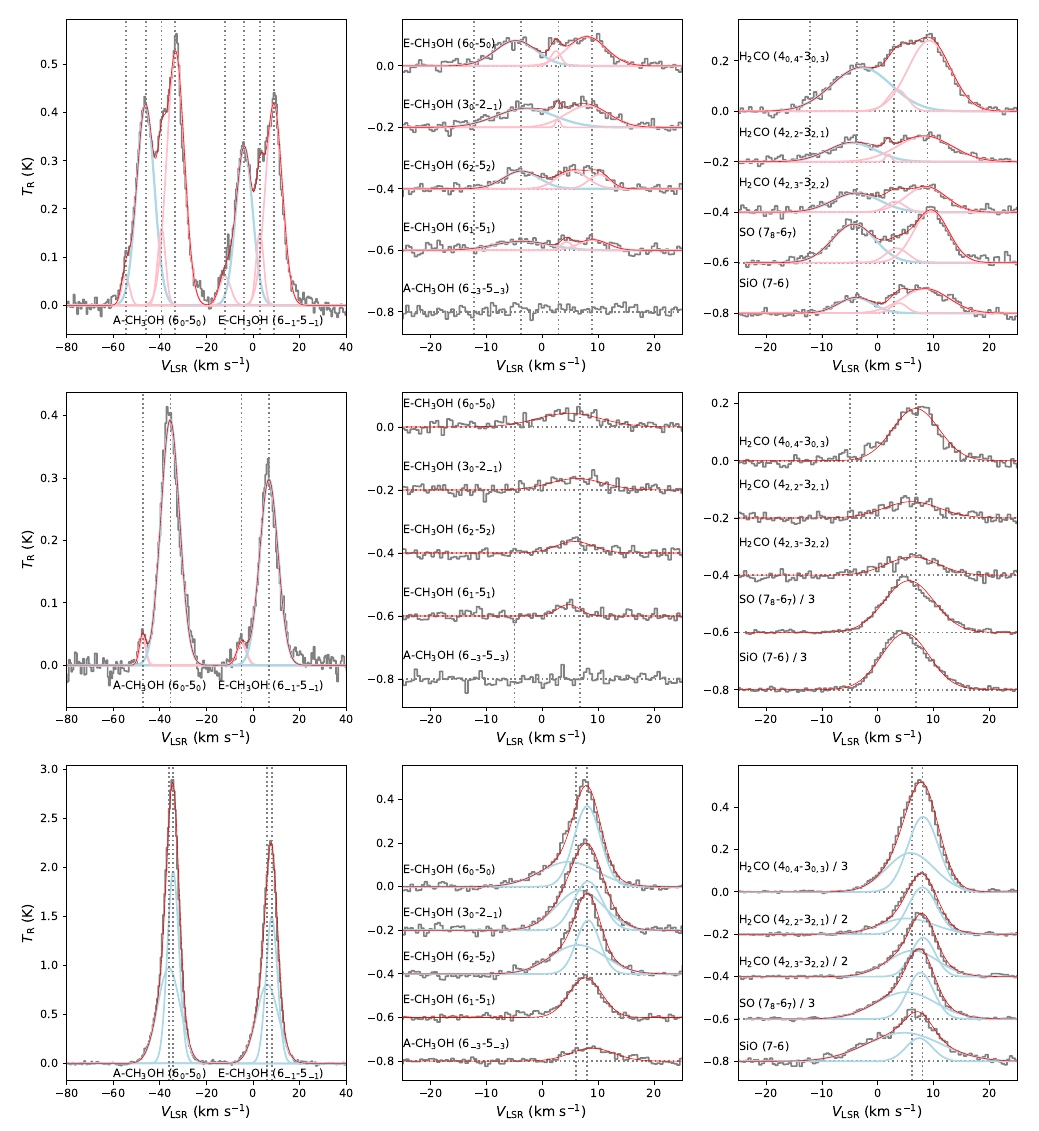}
\caption{Spectra (gray line) and the results of multi-Gaussian decomposition of all \emethanol, \amethanol, and \phhco\ lines, as well as the SO ($7_8$--$6_7$) and SiO (7--6) lines in regions 1--3 (from top to bottom). 
Some spectra are offset or scaled for better visualization. 
The red lines show the best-fit results. 
The light blue lines represent the component we mainly focus on, while the pink lines are the components that we do not discuss. 
The vertical dotted gray lines show the best-fit \vlsr\ of all components of the \emethanol\ line. 
\label{fig:fit123}
}
\end{figure*}

\begin{figure*}[!t]
\centering
\includegraphics[width=0.95\textwidth]{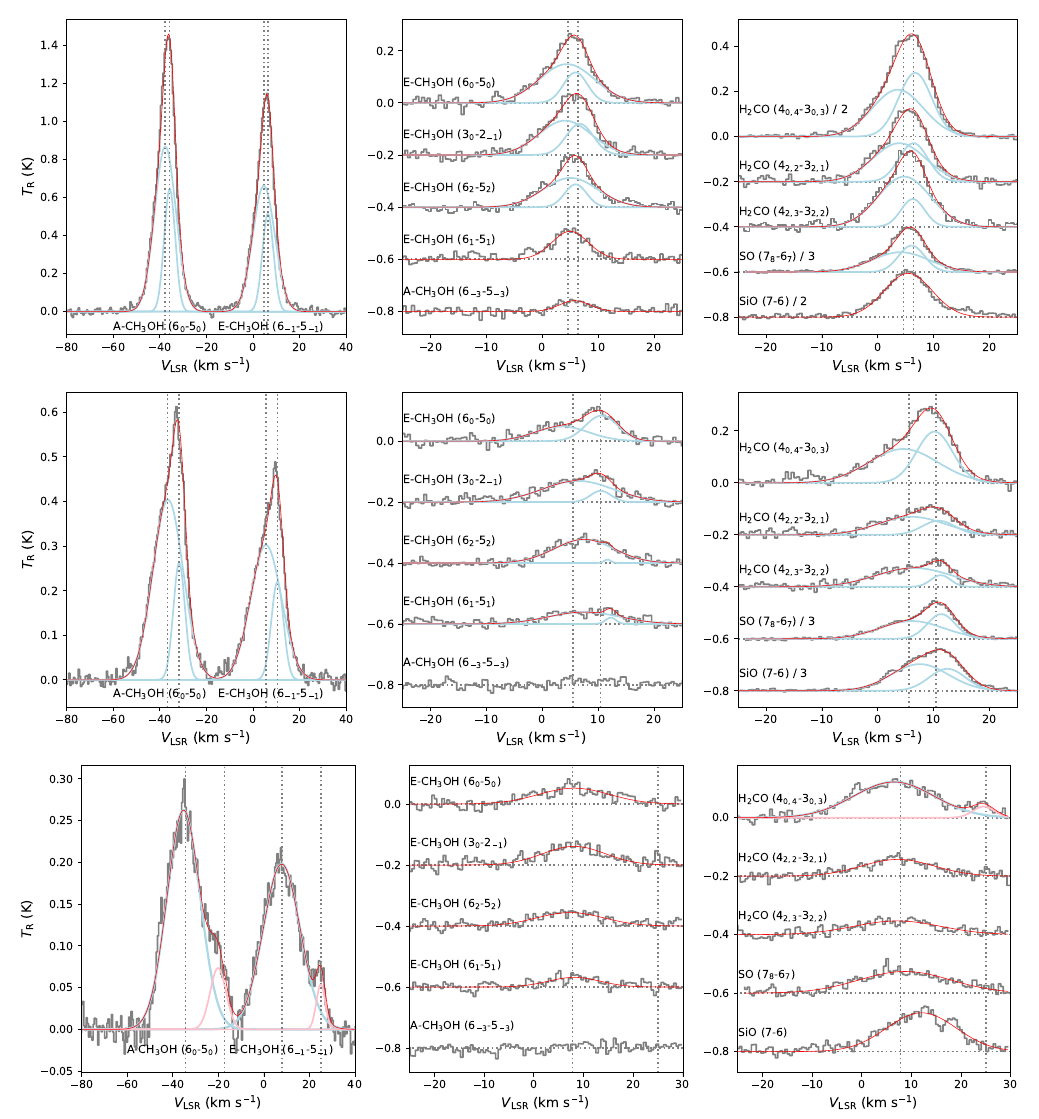}
\caption{Same as Figure \ref{fig:fit123} but for regions 4, 5, and 6. 
\label{fig:fit456}
}
\end{figure*}

To obtain the physical parameters of the shocked clumps, we first fit the detected transitions of \methanol, \phhco, SO, and SiO with one or multiple Gaussian components with \texttt{Python} package \texttt{lmfit}\footnote{\url{https://lmfit.github.io/lmfit-py/}}. 
The \coa\ and \cob\ (3--2) lines are not decomposed because  their profiles are complicated by unrelated line-of-sight components, multiple preshock and shocked components, and additional self-absorption in \coa\ (See Section \ref{sec:result_spec}). 
Decomposition of line profiles from shocked MCs into multiple Gaussian components has been conducted in previous studies \citep[e.g.,][]{Olmi_Herschel-HIFI_2015,Armijos-Abendano_Structure_2020,Hsieh_PRODIGE_2024a,Lin_Massive_2024a,Tu_Shock_2024}. 
However, we should always bear in mind that the line profiles of shocked MCs could be deviating significantly from the sum of several Gaussian components \citep[e.g.,][]{Tafalla_molecular_2010,Zhang_CO_2010,Lefloch_CHESS_2012}. 
Therefore, our spectral fitting should be regarded as an approximation. 
During the fitting, we first fit the \amethanol\ ($6_0$--$5_0$) and \emethanol\ ($6_{-1}$--$5_{-1}$) lines, and use the output $V_{\rm LSR}$ and FWHM as the input when fitting the other lines. 

\par

We fit the spectra of regions 3--5 with two Gaussian components.
These three regions exhibit significant non-Gaussian profiles, all with a blueshifted line wing structure probably brought by the shock interaction. 
Although line wing structures do not necessarily follow Gaussian lines profiles, we here fit the spectra toward these regions with two Gaussian components to separate the emission from the line center (high \vlsr\ component) and the line wing (low \vlsr\ component). 
We adopt three components in region 1 among which we mainly focus on the $\approx-3.8$ \kms\ component because its negative velocity sets it apart from other regions. 
Although the emission around $\sim +10$ \kms\ is stronger, it appears to be related to region 3 in both position and velocity, so we exclude it from our analysis of region 1.
For regions 2 and 6, we adopt two components for bright lines when necessary, and one component for the other lines. 
The results of the fitting are shown in Figure \ref{fig:fit123} and \ref{fig:fit456}, while the best-fit line parameters are listed in Table \ref{tab:fit}. 
The overall line profiles can be well characterized by our fitting. 
The SO and SiO lines in region 2 are not perfectly fitted by a single Gaussian component with the observed peaks slightly blueshifted from the best-fit peaks by $\approx0.7$ \kms. 

\par

We find that the line center \vlsr\ is often consistent among transitions of \emethanol, \amethanol, and \phhco, especially for the brightest lines \emethanol\ ($6_{-1}$--$5_{-1}$), \amethanol\ ($6_0$--$5_0$), and \phhco\ ($4_{0,4}$--$3_{0,3}$) (with high S/N ratios). 
Large discrepancies and uncertainties are found in faint transitions. 
In some cases, the \vlsr\ of SO and SiO deviate from the values of \emethanol\ ($6_{-1}$--$5_{-1}$) and \amethanol\ ($6_0$--$5_0$). 
In region 2 and the low \vlsr\ component of region 3, the SO and SiO lines are blueshifted, while in region 6 and both components of region 5, the SiO line is redshifted. 
Toward region 1, the \vlsr\ of the SiO and SO lines are consistent with those of other lines. 
In region 4, only one component of SiO (7--6) is found. 
The \vlsr\ of this component lies between the low and high \vlsr\ components. 
Its line FWHM is similar to that of the low \vlsr\ component, but there might be indistinguishable contribution from the high \vlsr\ component that affects the best-fit \vlsr. 

\par

The typical line FWHMs in all fitted components in regions 1, 2, and 6 are $\sim8.5$, $\sim 9$ and $\sim18$ \kms, respectively. 
It is possible that the fitted component in region 6 may actually consist of two separate velocity components, but we cannot distinguish them based on our data. 
For regions 3, 4, and 5 where two components are considered, the high \vlsr\ components tend to have line FWHM of $\approx5$ \kms, which also indicates shock origin, while the low \vlsr\ components tend to exhibit broader lines with $\rm FWHM\sim10$ \kms, which suggests stronger shock disturbance compared with the high \vlsr\ components. 
These broadened FWHMs are typical values in shocked dense MCs associated with SNRs \citep{vanDishoeck_Submillimeter_1993,Reach_Excitation_1999,Nicholas_12_2011,Tu_Shock_2024}. 
However, the best-fit line FWHM varies significantly among faint lines in the same region (e.g., regions 2 and 6). 
Analysis with the FWHM of faint lines with low signal-to-noise ratio should be cautious on the uncertainty of the fitted FWHM. 

\subsection{Estimation of physical parameters of the six shocked clumps}
\label{sec:nonLTE}

\begin{deluxetable*}{cccccccc}[!t]
\tablecaption{Derived physical parameters of fitted components of the six molecular clumps based on the MCMC and non-LTE analysis. 
\label{tab:param}}
\tablehead{ 
\colhead{Region} & 
\colhead{\makecell{$n_{\rm H_2}$$^{\rm a}$ \\$\rm \ (10^5\ cm^{-3})$} } & 
\colhead{\makecell{$T_{\rm gas}$$^{\rm a}$ \\(K)}} & 
\colhead{\makecell{$N$(\emethanol)$^{\rm a}$ \\$\rm (10^{13}\ cm^{-2})$}} & 
\colhead{\makecell{$N$(\amethanol)$^{\rm a}$ \\$\rm (10^{13}\ cm^{-2})$}} & 
\colhead{\makecell{$N$(\phhco)$^{\rm a}$ \\$\rm (10^{13}\ cm^{-2})$}} & 
\colhead{\makecell{$N$({\rm SO})$^{\rm a}$ \\$\rm (10^{13}\ cm^{-2})$}} & 
\colhead{\makecell{$N$({\rm SiO})$^{\rm a}$ \\$\rm (10^{13}\ cm^{-2})$}}
}
\startdata 
1 & $1.7^{+0.6}_{-0.4}$ & $86^{+17}_{-13}$ & $7.5^{+1.1}_{-1.0}$ & $6.1^{+0.8}_{-0.8}$ & $1.6^{+0.4}_{-0.3}$ & $5.5^{+1.8}_{-1.4}$ & $0.68^{+0.28}_{-0.21}$ \\ 
2 & $2.0^{+0.6}_{-0.4}$ & $106^{+24}_{-18}$ & $5.3^{+0.5}_{-0.4}$ & $5.6^{+0.6}_{-0.6}$ & $1.0^{+0.2}_{-0.2}$ & $15.0^{+3.4}_{-2.8}$ & $4.8^{+1.2}_{-1.0}$ \\ 
3 &  &  &  &  &  &  &  \\ 
Low $V_{\rm LSR}$ & $1.5^{+1.4}_{-0.6}$ & $107^{+31}_{-22}$ & $17.4^{+5.9}_{-4.5}$ & $17.9^{+4.0}_{-3.6}$ & $3.9^{+1.9}_{-1.5}$ & $16.2^{+9.1}_{-6.5}$ & $2.6^{+1.9}_{-1.4}$ \\ 
High $V_{\rm LSR}$ & $2.4^{+0.8}_{-0.6}$ & $86^{+19}_{-13}$ & $18.3^{+2.4}_{-2.2}$ & $18.9^{+2.6}_{-2.5}$ & $3.2^{+0.7}_{-0.6}$ & $9.9^{+2.6}_{-2.1}$ & $0.58^{+0.22}_{-0.17}$\\ 
4 &  &  &  &  &  &  &  \\ 
Low $V_{\rm LSR}$ & $2.0^{+1.0}_{-0.6}$ & $145^{+70}_{-38}$ & $15.8^{+2.2}_{-1.9}$ & $13.6^{+2.2}_{-2.0}$ & $2.7^{+0.9}_{-0.6}$ & $6.8^{+2.6}_{-1.8}$ & $2.6^{+1.2}_{-0.9}$ \\ 
High $V_{\rm LSR}$ & $2.9^{+4.1}_{-1.4}$ & $51^{+16}_{-11}$ & $6.8^{+1.4}_{-1.2}$ & $6.7^{+1.5}_{-1.4}$ & $1.7^{+0.8}_{-0.6}$ & $7.6^{+4.1}_{-2.9}$ & ---$^{\rm b}$ \\ 
5 &  &  &  &  &  &  &  \\ 
Low $V_{\rm LSR}$ & $1.3^{+0.3}_{-0.3}$ & $162^{+56}_{-32}$ & $11.2^{+1.2}_{-1.1}$ & $9.0^{+1.0}_{-1.0}$ & $1.8^{+0.4}_{-0.3}$ & $7.6^{+2.0}_{-1.6}$ & $3.4^{+1.3}_{-1.0}$ \\ 
High $V_{\rm LSR}$ & $16.4^{+24.8}_{-9.4}$ & $62^{+43}_{-17}$ & $3.2^{+0.7}_{-0.5}$ & $3.4^{+1.2}_{-0.7}$ & $0.20^{+0.06}_{-0.04}$ & $2.3^{+0.6}_{-0.4}$ & $0.15^{+0.14}_{-0.05}$ \\ 
6 & $1.4^{+0.3}_{-0.3}$ & $113^{+22}_{-17}$ & $9.7^{+1.1}_{-1.0}$ & $7.1^{+0.8}_{-0.8}$ & $2.0^{+0.4}_{-0.3}$ & $6.1^{+1.6}_{-1.3}$ & $3.1^{+0.9}_{-0.7}$\\ 
\hline
\enddata
\tablecomments{
$^{\rm a}$ The superscripts and subscripts denote the $1\sigma$ uncertainties of the fitted physical parameters. 
$^{\rm b}$ The fitting of SiO in region 4 only yield one component in the low \vlsr\ component. It is unclear whether there is contribution from the high \vlsr\ component.
}
\end{deluxetable*}

\begin{deluxetable}{cccc}[ht]
\tablecaption{Estimated column densities ratios. 
\label{tab:SOSiO}}
\tablehead{ 
\colhead{Region} & 
\colhead{E/A ratio$^{\rm a}$} & 
\colhead{$\frac{N({\rm SO})}{N({\rm CH_3OH})}$} & 
\colhead{$\frac{N({\rm SiO})}{N({\rm CH_3OH})}$}  
}
\startdata 
1 & $1.2^{+0.2}_{-0.2}$ & $0.41^{+0.09}_{-0.08}$ & $0.050^{+0.016}_{-0.013}$  \\ 
2 & $0.94^{+0.12}_{-0.11}$ & $1.4^{+0.3}_{-0.3}$ & $0.44^{+0.10}_{-0.09}$  \\ 
3 &  & &   \\ 
Low \vlsr & $1.0^{+0.3}_{-0.3}$ & $0.46^{+0.15}_{-0.14}$ & $0.076^{+0.034}_{-0.034}$  \\ 
High \vlsr & $0.97^{+0.20}_{-0.16}$ & $0.27^{+0.05}_{-0.05}$ & $0.016^{+0.005}_{-0.005}$  \\ 
4 & & &   \\ 
Low \vlsr & $1.2^{+0.2}_{-0.2}$ & $0.23^{+0.07}_{-0.06}$ & $0.089^{+0.035}_{-0.029}$  \\ 
High \vlsr & $1.0^{+0.4}_{-0.2}$ & $0.58^{+0.23}_{-0.20}$ & ---$^{\rm b}$  \\ 
5 & & &   \\ 
Low \vlsr & $1.2^{+0.2}_{-0.1}$ & $0.49^{+0.12}_{-0.11}$ & $0.23^{+0.05}_{-0.05}$  \\ 
High \vlsr & $0.91^{+0.22}_{-0.19}$ & $0.33^{+0.18}_{-0.09}$ & $0.022^{+0.029}_{-0.010}$  \\ 
6 & $1.4^{+0.2}_{-0.2}$ & $0.36^{+0.07}_{-0.06}$ & $0.18^{+0.05}_{-0.03}$ \\ 
\hline
\enddata
\tablecomments{
$^{\rm a}$ $N(\text{E-}{\rm CH_3OH})/N(\text{A-}\rm CH_3OH)$.  
$^{\rm b}$ $N(\rm SiO)$ is not obtained in the high \vlsr\ component of region 4 (See Table \ref{tab:param}).}
\end{deluxetable}

After the spectral decomposition, we can estimate the physical parameters of each component. 
The detected multiple transitions of \emethanol\ and \phhco\ allow us to conduct non-local-thermodynamic-equilibrium (non-LTE) analysis with the RADEX radiative transfer code \citep{vanderTak_computer_2007}. 
The plane-parallel geometry is chosen which is applicable to shocked MCs, with the escape probability $\beta=(1-\exp{(-3\tau)})/3\tau$ where $\tau$ is the optical depth. 
The collisional rate coefficients of the considered species are taken from the EMAA database, with those of \emethanol\ from \citet{Dagdigian_Rotational_2024} and those of \phhco\ from \citet{Wiesenfeld_Rotational_2013}. 
The \texttt{Python} package \texttt{emcee} \citep{Foreman-Mackey_emcee_2013} is then used to carry out Markov Chain Monte Carlo (MCMC) sampling of the values to be fitted (i.e. \hh\ volume density $n_{\rm H_2}$, gas temperature $T_{\rm gas}$, and molecular column densities $N(\rm species)$) with uniform distribution for priors.
In our first attempt, the fitting of the \phhco\ or \emethanol\ lines independently does not converge. 
This is probably because the critical densities of all the \phhco\ lines are similar which causes that the predicted integrated intensities of the \phhco\ lines depends poorly on the \hh\ density. 
For \emethanol, although the critical density of the ($3_0$--$2_{-1}$) line is much higher than those of the other lines, they still put poor constraints on the \hh\ density. 
To make the model converge, we choose to fit all the detected \phhco\ lines, \emethanol\ lines, and the \amethanol\ ($6_0$--$5_0$) lines simultaneously. 
The \amethanol\ ($6_{-3}$--$5_{-3}$) line is not fitted because it is only detected in regions 3 and 4. 
By fitting all of the transitions at the same time, we assume that they trace the molecular gas with similar physical conditions, i.e. the same $n_{\rm H_2}$ and $T_{\rm gas}$. 
This is a reliable assumption for \methanol\ and \phhco\ because all these transitions show similar spatial distribution and spectral components. 
The uncertainty of the observed integrated intensities is estimated as the sum of the fitting uncertainty and the assumed uncertainty in flux calibration, 10\% of the total flux, which is slightly higher than the typical value of 5\% (see the ALMA Cycle 12 Technical Handbook\footnote{\url{https://almascience.nrao.edu/documents-and-tools/cycle12/alma-technical-handbook}}). 
Throughout our non-LTE fitting, the column densities of \phhco, \emethanol, and \amethanol\ are independent parameters, and we do not consider the problem of the unknown beam filling factor. 
The input linewidth is fixed to the average FWHM of all the fitted transitions in the component.
After obtaining the best-fit $n_{\rm H_2}$ and $T_{\rm gas}$, we use their posterior distributions to estimate the column densities of SiO and SO with the SiO (7--6) and SO ($7_8$--$6_7$) lines, with the collisional rate coefficients obtained from \citet{Balanca_Rotationally_2018} and \citet{Lique_Rotationally_2007}, respectively. 
This is based on the assumption that the physical parameters of the gas traced by SiO and SO are comparable to those of \phhco\ and \methanol.
Although this assumption may not hold true, this is the only way that we can manage to obtain an estimate the column densities of SiO and SO for reference.

\par

The best-fit physical parameters and their uncertainties obtained from the MCMC and non-LTE analysis are shown in Table \ref{tab:param}. 
The MCMC process converges in all components. 
But in the high \vlsr\ component of region 5, the uncertainty in $n_{\rm H_2}$ is very high. 
We note that we do not use the \emethanol\ ($6_2$--$5_2$) and ($6_1$--$5_1$) in the fitting of the high \vlsr\ component of region 5 because they are very faint (see Figure \ref{fig:fit456}). 
The lack of these two lines may affect the results of the fitting. 
The reason why region 3 shows the brightest \methanol\ emission is likely to be simply due to the high column density instead of a favorable excitation condition. 

\par

In all components except the high \vlsr\ component of region 5, the best-fit \hh\ density is around $(1\text{--}3)\times10^5\rm \ cm^{-3}$, exhibiting very moderate density fluctuation. 
This range is lower than the best-fit value found by \citet{Mazumdar_Submillimeter_2022a}, which is $(0.9\text{--}5)\times 10^6\rm \ cm^{-3}$. 
This could be because \citet{Mazumdar_Submillimeter_2022a} included some \hhco\ lines with higher critical densities (up to $\sim 10^8\rm \ cm^{-3}$) than those of the lines we detect, so the transitions they observed may trace denser parts of the shocked MC. 
The best-fit gas temperatures vary from $\sim 50$ K to $\sim 170$ K, which are typical values found in shocked MCs probed by various molecular tracers \citep{vanDishoeck_Submillimeter_1993,Maxted_Ammonia_2016}. 

\begin{figure}[!t]
\centering
\includegraphics[width=0.47\textwidth]{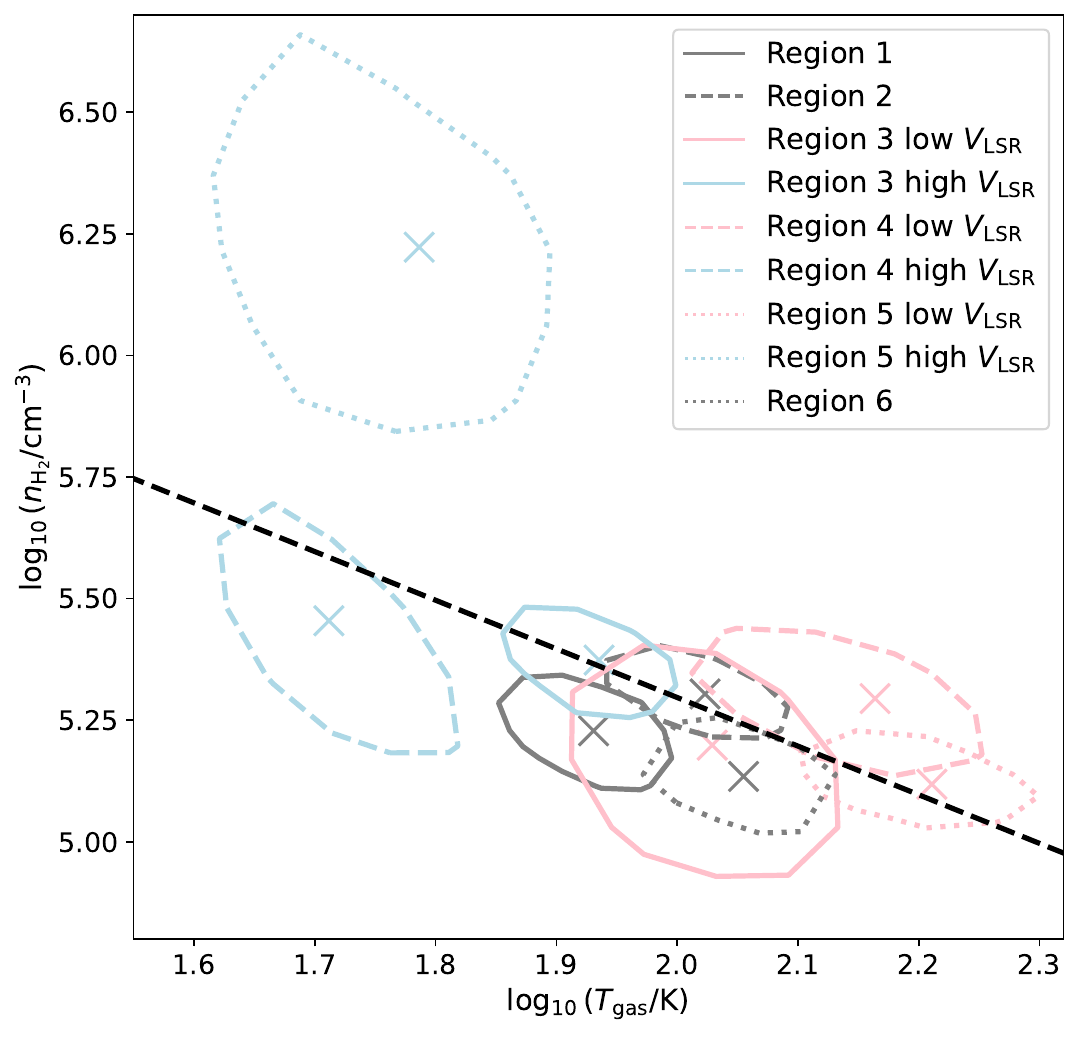}
\caption{A scatter plot between the best-fit gas temperature ($T_{\rm gas}$) and \hh\ density ($n_{\rm H_2}$) in Table \ref{tab:param} shown in crosses, overlaid with contours showing their $1\sigma$ confidence regions. 
The dashed black line shows the best-fit result of a constant $n_{\rm H_2}T_{\rm gas}$ except for the high \vlsr\ component of region 5. 
\label{fig:nT}
}
\end{figure}

In Figure \ref{fig:nT}, we make a scatter plot of the best-fit gas temperature and \hh\ density as well as their $1\sigma$ confidence regions. 
These two parameters clearly show anti-correlation. 
In regions 3--5, the low \vlsr\ component always has lower \hh\ density and higher gas temperature than the high \vlsr\ component. 
The thermal pressure of MC can be written as $P_{\rm gas}/k_{\rm b}\sim n_{\rm H_2}T_{\rm gas}$. 
By fitting a constant $P_{\rm gas}$ in Figure \ref{fig:nT} except for the high \vlsr\ component of region 5 which is an outlier, we find that all the six regions tend to have a typical pressure of $\sim 2 \times 10^7\rm \ cm^{-3}\ K$. 
On the other hand, the thermal pressure of the X-ray ear (see Figure \ref{fig1}), which is located close to W28F, is $P_{\rm X}/k_{\rm b}\sim 2.3n_{\rm H}T_{\rm X}$, where $n_{\rm H}$ and $T_{\rm X}$ are the hydrogen density and the temperature of the X-ray emitting gas. 
According to \citet{Zhou_XMM-Newton_2014}, we obtain $n_{\rm H}\sim 2.5\rm \ cm^{-3}$ and $k_{\rm b}T_{\rm X}\sim 0.3\rm \ keV$. 
The thermal pressure of the X-ray emitting gas is then $\sim 2 \times 10^7\rm \ cm^{-3}\ K$. 
The consistency in the thermal pressure between the shocked molecular gas and the X-ray emitting gas strongly suggests the physical association between these two phases of the shocked ISM. 
Turbulence may cause extra pressure in the shocked molecular gas, but its contribution is hard to quantify. 
Further study is needed to constrain the turbulence pressure in such complicated environment.
We also note from Figure \ref{fig:nT} that resulting sample of $n_{\rm H_2}$ and $T_{\rm gas}$ in each single component are moderately anti-correlated, with Pearson correlation coefficients varying from $\approx -0.4 $ to $\approx-0.5$ except the low \vlsr\ component of region 3 ($-0.24$) and the high \vlsr\ component of region 5 ($-0.69$). 
So the correlation between $n_{\rm H_2}$ and $T_{\rm gas}$ may be partly due to the degeneracy in the fitting in addition to the scenario of constant pressure. 
Observation of more transitions may help reduce the degeneracy and verify the scenario of constant pressure. 

\par
In addition to the dense molecular clumps and the hot ionized gas, we can also estimate the pressure of the extended warm (100--4000 K) \hh\ gas with the pure rotational lines of \hh\ observed by \citet{Yuan_Spitzer_2011}. 
The temperature distribution of the warm \hh\ gas follows a power law, with the differential density of gas at temperature between $T$ and $T + dT$ proportional to $T^{-3.36}$. 
Provided a total density of $10^{3.83}\rm \ cm^{-3}$, the estimated pressure of the warm \hh\ gas is $10^6 \rm \ cm^{-3}\ K$, which is approximately one order of magnitude lower than the dense molecular clumps and the hot ionized gas. 
This could be because a large fraction of the total pressure in warm \hh\ gas is present as turbulent pressure instead of thermal pressure \citep{Krumholz_Star_2009}. 
The typical line FWHMs of the far-infrared (FIR, 100--190 $\mu$m) CO rotational lines observed by Herschel are $\sim 200$ \kms\ \citep{Neufeld_Water_2014}\footnote{This FWHM was roughly estimated based on Figure 7 and Table 2 of \citet{Neufeld_Water_2014}. For example, their Table 2 shows that the integrated flux of the CO $J=18$--17 line at 144.78 $\mu$m toward W28F is $1.46\times10^{-16}\rm \ W\ m^{-2}$. Its peak flux density can be estimated from their Figure 4 to be $\sim 7$ Jy. Assuming a Gaussian line profile, the FWHM can be estimated to be $\sim 280$ \kms. Note that the typical spectral resolution adopted by \citet{Neufeld_Water_2014} is $\sim 1500$, i.e. $\sim 200$ \kms, at this wavelength. Therefore, the FWHM of the CO line, even if we remove the contribution from the instrument, is $\sim 200$ \kms. Such a large line width can be caused by the turbulent motion, the unresolved velocity gradients \citep{Reach_Supernova_2019,Lehmann_Self-generated_2022}, as well as the uncertainty in the instrumental broadening.}, which is higher than the typical linewidth of the molecular lines we detected by an order of magnitude. 
Provided that the FIR CO lines trace a similar physical condition as the \hh\ pure rotational lines, such large linewidths suggest that the turbulence in the warm \hh\ gas is significantly stronger than in the dense shocked molecular clumps traced by \methanol\ and SiO, although the large linewidths may also include contribution from some unresolved velocity structures. 
Future infrared \hh\ observation with higher angular and spectral resolution will help explain the low thermal pressure associated with W28F.

\subsection{The E/A ratio of \methanol} \label{sec:E/A}
The \methanol\ molecule has two spin isomers, \emethanol\ and \amethanol. 
The radiative or collisional transition between \emethanol\ and \amethanol\ is forbidden, and they are thus often treated as two separate species \citep{Wirstrom_Observational_2011}. 
Since the conversion between the two isomers requires a long timescale, the E/A ratio of \methanol\ is expected to be fixed ever since it is formed. 

\par

The ground state of \emethanol\ is 7.9 K higher than that of \amethanol. 
Therefore, if the formation environment (provided that it is in thermal equilibrium) of \methanol\ is at a low temperature, we expect that the E/A ratio of \methanol\ to be $<1$. 
On the contrary, if the formation environment is warm enough \citep[$\gtrsim 37$ K,][]{Wirstrom_Observational_2011}, the E/A ratio can be $\sim 1$. 
An important formation mechanism of \methanol\ is successive hydrogenation of CO ice \citep[i.e., $\rm CH_3O/CH_2OH+H$,][]{Watanabe_Hydrogenation_2004,Fuchs_Hydrogenation_2009} or reaction $\rm CH_3O+H_2CO$ \citep{Simons_Formation_2020} on dust mantle at typical temperature of quiescent dark MCs ($\sim 15$ K), followed by desorption to the gas phase, such as thermal desorption \citep{Bisschop_Testing_2007}, shock sputtering \citep{Bachiller_Methanol_1995} and other various processes \citep[e.g.,][]{Wakelam_Efficiency_2021}. 
If this is the case, $\rm E/A<1$ is expected because \methanol\ forms on cold ($\sim 15$ K) dust mantles. 
Such values have been found in cold dense MCs and protostellar outflows where the formation of gas-phase \methanol\ is expected to follow this route \citep[e.g.,][]{Menten_Methanol_1988,Holdship_Observations_2019}. 
However, it is also common to find E/A close to or even higher than 1 \citep{Purcell_Physical_2009}. 

\par

As shown in Table \ref{tab:param}, the \methanol\ E/A ratio in the six regions are all higher than 0.9, and in some cases higher than 1, e.g. regions 1 and 6, as well as the low \vlsr\ components of regions 4 and 5. 
If the gas-phase transformation between \emethanol\ and \amethanol\ can be neglected, $\rm E/A=0.9$ will require a formation temperature of $>16$ K \citep{Wirstrom_Observational_2011}. 
At this temperature, \methanol\ still forms efficiently on dust mantle. 
However, it is not clear whether methanol formation on dust grain surfaces alone can explain even higher E/A ratios that we detect ($> 0.9$) since it would require higher dust temperatures. 
The preshock MC appears in \coa\ self-absorption in W28F (see Figure \ref{fig:COspec} and Section \ref{sec:result_spec}), which indicates that the preshock gas is at very low temperature (at least lower than the brightness temperature of the \coa\ 3-2, for instance, $\sim16$ K for regions 1, 2, 5, and 6). 
Simulations proposed that in shocked MCs, \methanol\ forms mainly via desorption from dust grains instead of gas-phase reactions \citep[e.g.,][]{Huang_Investigating_2025b}. 
Therefore, additional mechanisms are required to explain the high E/A ratio.

\par

\citet{Wirstrom_Observational_2011} proposed that proton exchange reactions with $\rm H_3^+$ and $\rm HCO^+$ can equilibrate the E/A ratio at another characteristic temperature. 
Since enhanced CR ionization rates have been found in W28 \citep{Vaupre_Cosmic_2014,Indriolo_Absorption-line_2023,Tu_Shock_2024}, it is possible that $\rm H_3^+$, a direct product of CR ionization, plays a role in raising the E/A ratio. 
Such mechanism has been used to explain the high E/A in the central molecular zone of NGC\,253 \citep{Huang_Investigating_2025} where enhanced CR ionization rates are also found \citep{Harada_Starburst_2021,Holdship_Energizing_2022}. 
In addition, the assumptions and uncertainties in the non-LTE analysis, as well as some excitation effects \citep{Hernandez-Hernandez_APEX_2019}, may also lead to the enhanced E/A ratio that we infer from observation.
We note that the enhanced E/A ratio is expected to be the result of an interplay of multiple processes, which cannot be readily distinguished based on the current observation.

\subsection{Chemical segregation between \methanol\ and SiO} \label{sec:compare_CH3OH_SiO}
Both SiO and \methanol\ have been regarded as shock tracers. 
SiO is believed to form via the sputtering of dust grain cores in fast shocks ($\gtrsim25$ \kms), followed by gas phase reactions of $\rm Si+O_2/OH$ \citep{Schilke_SiO_1997,Caselli_Grain-grain_1997,Gusdorf_SiO_2008}. 
In addition to sputtering, processes such as grain-grain collision \citep{Caselli_Grain-grain_1997}, grain vaporization \citep{Guillet_Shocks_2011}, and desorption of ice mantle with pre-existing SiO \citep{Nguyen-Luong_Low-velocity_2013}, can also contribute to the total yield of gas-phase SiO when the shock velocity is lower \citep[$\gtrsim15$--20 \kms,][]{Caselli_Grain-grain_1997,Jimenez-Serra_Parametrization_2008}. 
But in general, more SiO is expected to be produced in MCs subject to faster shocks because they can release more Si atoms to the gas phase from dust grains \citep{Garay_Silicon_2000,Gusdorf_SiO_2008,Hiramatsu_Outflow-Core_2010a}. 
On the other hand, \methanol\ mainly resides in dust mantles and are desorbed to the gas phase in shocked MCs (see Section \ref{sec:E/A}). 
Simulations found that gas-phase chemistry can produce a \methanol\ abundance up to $\sim10^{-8}$, while the inclusion of mantle desorption will raise this value to $\sim 10^{-7}$--$10^{-6}$ \citep{Bergin_Postshock_1998,Jimenez-Serra_Parametrization_2008}. 
The sputtering and evaporation of dust grain mantles requires much slower shocks than the sputtering of grain cores to produce SiO. 
A shock with a velocity of $\sim 10$ \kms\ is already fast enough to enhance the \methanol\ abundance compared with quiescent MCs \citep{Jimenez-Serra_Parametrization_2008,Huang_Tracing_2023}. 
In fast shocks, however, some studies found \methanol\ can be destroyed during the sputtering or by the following gas-phase reactions \citep{Jorgensen_impact_2004,Suutarinen_Water_2014}, while some found high \methanol\ abundance even after the passage of a shock of $\sim40$--50 \kms\ \citep{Jimenez-Serra_Parametrization_2008,Huang_Tracing_2023}. 

\par

Our observation reveals the chemical segregation between \methanol\ and SiO in both the spatial and spectral points of view. 
Their spatial distribution is different, as shown in Figure \ref{fig:integrated_intensity_1} and \ref{fig:channel_maps}, with \methanol\ peaking at region 3 and SiO peaking at regions 2 and 5. 
Details about the different spatial and spectral distribution between \methanol\ and SiO have been presented in Section \ref{sec:integrated_intensity} and \ref{sec:velocity_channel_map}. 
The estimation of the column density of SiO further confirm the enhanced \sioonmethanol\ abundance ratio in region 2 and the low \vlsr\ component of region 5 (see Table \ref{tab:SOSiO}). 
In addition, the \sioonmethanol\ in the low \vlsr\ components are always higher than the values in the high \vlsr\ components of regions 3--5, where the line widths are smaller and \hh\ densities are higher. 

\par

All of these observational findings suggest the scenario that \methanol\ formation on dust grain surface followed by desorption in low-velocity shocks should be considered as the dominant formation mechanism of gas-phase \methanol, while SiO traces the fast shocks. 
This scenario can explain both the spatial and spectral features of the \methanol\ and SiO emission. 

\par
 
The smaller densities and the larger line widths in the low \vlsr\ components, where the \sioonmethanol\ abundance ratios are high, can both be attributed to faster shocks. 
When an SNR blast wave propagates into a MC with density fluctuation, the cloud shocks in the more diffuse clouds is expected to be faster than those in the denser clouds, and can drive stronger disturbance which gives rise to larger line widths \citep{Rho_Detection_2015,Reach_Supernova_2024}. 
Toward region 2, i.e. the SiO peak, the blueshift of the SiO line can also be explained by faster shocks pushing the SiO line farther away from the systemic velocity of W28. 
We note that the single Gaussian fitting of the line profile at region 2 actually underestimates the extent of this blueshift because the line profile of SiO deviates from a standard Gaussian profile (see Figure \ref{fig:integrated_intensity_1}). 
Enhanced line ratio between SiO and \methanol\ at line wings has also been observed frequently in protostellar outflows, where SiO emission is often found to be enhanced in the high velocity components \citep{Jorgensen_impact_2004,Tafalla_molecular_2010,Codella_SOLIS_2025a}. 

\par

Typical values of \sioonmethanol\ in Galactic protostellar outflows fall within $\sim3\times10^{-4}$--$4\times 10^{-2}$ \citep[e.g.,][]{Garay_Molecular_1998,Garay_Methanol_2002,Jorgensen_impact_2004,Tafalla_HH_2013,Sabatini_FAUST_2024,Codella_SOLIS_2025a}. 
In comparison, we detect $N({\rm SiO})/N({\rm CH_3OH})\gtrsim0.1$ toward three regions (components), i.e. region 2 (0.44), region 6 (0.23), and the low \vlsr\ component of region 5 (0.17), higher than typical values in protostellar outflows. 
Such high value of \sioonmethanol\ has been observed in the extremely high-velocity component of protostars \citep[e.g.,][]{Tafalla_molecular_2010}. 
This further suggests that the segregation between \methanol\ and SiO in W28F is due to different shock velocities. 
However, we should still bear in mind that here the SiO column density is estimated based on the assumption that it traces the same gas as \methanol, which is likely not the case. 
Since the molecular gas traced by SiO can be subject to faster shock than that traced by \methanol, SiO may reside in clumps with higher temperature and lower density than those traced by \methanol. 
In this case, the column density of SiO we obtain may be overestimated. 
We also note that obtaining the abundances of \methanol\ and SiO relative to \hh\ and comparing them to typical values in other environments would be a good way to study their origin in W28F. 
However, we cannot determine $N(\rm H_2)$ toward W28F since the CO lines are hard to decompose, and even if the column density of CO is estimated, the abundance of CO could be significantly lower than the typical interstellar values due to shock destruction \citep[e.g.,][]{Zhou_Unusually_2022b}. 

\subsection{Comparison between \methanol\ and other species}

As shown in Figures \ref{fig:integrated_intensity_2}, \ref{fig:fit123}, and \ref{fig:fit456}, the spatial distribution and spectral features of \methanol\ and \phhco\ lines are rather similar. 
This suggests that the chemistry of these two molecular species is similar to each other, both formed on ice and desorbed by the shock wave \citep{Cuppen_Microscopic_2009,Gomez-Ruiz_1157_2013,Burkhardt_Modeling_2019}. 
SO, however, peaks at region 3 but also show clumpy structures at regions 2 and 5. 
The \soonmethanol\ is highest at region 2 ($\approx 1.4$), and drops to 0.2--0.6 in several other components. 
The SO distribution is in an intermediate manner between \methanol\ and SiO suggests that it traces the intermediate velocity shocks. 
Some studies also proposed the similar opinion \citep[e.g.,][]{Codella_SOLIS_2025a}.
Indeed, SO is a species widely found in both cold quiescent MCs and shocks with low and high velocities \citep{Tychoniec_Which_2021,Fuente_Gas_2023,Bouvier_ALCHEMI_2024}. 
Enhancement of SO abundance in shocked MCs could be due to the desorption of S-bearing species, such as atomic S and $\rm H_2S$, from the dust mantle, followed by gas-phase reaction with O, $\rm O_2$ and OH \citep{PineaudesForets_Sulphur-bearing_1993,Tychoniec_Which_2021}. 

\par

The column densities of CO in this study are not estimated due to their complex line profiles. 
Figure \ref{fig:COspec} shows that the line profiles of \cob\ and \amethanol\ are similar in the red wings but not in the blue wings. 
According to our discussion in Section \ref{sec:compare_CH3OH_SiO}, the red wing is likely due to the low-velocity shock, while the blue wing is associated with the high-velocity shock that produces SiO. 
In shocked MCs, the abundance of CO is not expected to be enhanced or reduced significantly compared to its canonical value without depletion ($\sim 10^{-4}$) in quiescent MC \citep{Flower_Excitation_2010}. 
The similar line profiles of \cob\ and \amethanol\ in the red wings suggest their similar kinematics under the shock interaction, although they may trace different parts of the MC due to their different critical densities. 
However, in fast shocks ($\gtrsim 25$ \kms), the CO molecules could be dissociated \citep{Zhou_Unusually_2022b}. 
The different line profiles between \cob\ and \amethanol\ in the blue wings could thus be due to their different chemistry in fast shocks. 
Alternatively, it could also arise from their different dependency on the drastic density enhancement by the fast shock, provided their distinct critical densities. 

\section{Conclusion} \label{sec:con}

We present new ACA observation of W28F, an MC interacting with SNR W28 featuring the detection of five 1720 MHz OH masers and two 36 GHz \methanol\ masers. 
We detect CO isotopes, \methanol, \phhco, SO, $\rm ^{34}SO$, SiO, and OCS lines. 
Our main findings are summarized as follows. 

1. Clumpy structures of shocked MCs are revealed by our observation, with masers distributed roughly at their outer edge. 
All \methanol\ and \phhco\ lines show similar spatial distribution, different from that of SO and SiO. 

2. We choose six shocked clumps from the velocity channel maps of \amethanol\ and SiO and conduct (multi-)Gaussian spectral decomposition as well as non-LTE radiative transfer analysis in these regions. 
The brightest \methanol\ and \phhco\ lines always show consistent spectral line profiles, while the line centers of the SiO line sometimes deviate from their \methanol\ counterparts. 
The best-fit \hh\ density and gas temperature in most of the fitted components fall in 1--$3\times10^5\rm \ cm^{-3}$ and 50--160 K, respectively. 
We find an anti-correlation between $n_{\rm H_2}$ and $T_{\rm gas}$, from which we estimate an average thermal pressure of $\sim 2\times 10^7\rm \ cm^{-3}\ K$, which is consistent with the thermal pressure of the X-ray emitting hot plasma close to W28F. 
This is consistent with the picture that the SNR shocks propagate into multi-phase gas, with a pressure balance existing between different phases.

3. The E/A ratios of \methanol\ in all the fitted components are $\gtrsim 0.9$, which suggests some extra processes of \methanol\ in the gas phase, such as proton exchange with $\rm H_3^+$ and \hcop. 
The chemical segregation between \methanol\ and SiO in both spatial distribution and spectral line profile is likely because \methanol\ traces slower shocks ($\sim10$ \kms) than SiO does ($\gtrsim 25$ \kms). 
The chemistry of \methanol\ and \phhco\ is expected to be similar because their behavior in W28F is similar. 
SO is likely to trace intermediate velocity shocks.

\begin{acknowledgments}
This work is partially supported by National SKA Program of China (2025SKA0140100). 
T.-Y. T. acknowledges the financial support of the China Scholarship Council (No. 202406190190). 
T.-Y. T. thanks Pei-Ying Hsieh for helpful discussions on observation and data reduction, and Yi-Heng Chi for providing the X-ray EP-FXT image of W28. 
W.Y. acknowledges the support from NSFC (12403027), China Postdoctoral Science Foundation (2024M751376), and Jiangsu Funding Program for Excellent Postdoctoral Talent (2024ZB347).
S.F. acknowledges the support from the National Key R\&D program of China grant (2025YFE0108200) and NSFC grant No. 12373023. 
V.W. acknowledges the CNRS program ``Physique et Chimie du Milieu Interstellaire'' (PCMI) co-funded by the Centre National d’Etudes Spatiales (CNES).
Y.C. acknowledges the support from NSFC grants Nos. 12573047, 12121003 and 12393852. 
P.Z. acknowledges the support from NSFC grant No. 12273010.

\par

This paper makes use of the following ALMA data: ADS/JAO.ALMA\#2024.1.00194.S. ALMA is a partnership of ESO (representing its member states), NSF (USA) and NINS (Japan), together with NRC (Canada), NSTC and ASIAA (Taiwan), and KASI (Republic of Korea), in cooperation with the Republic of Chile. The Joint ALMA Observatory is operated by ESO, AUI/NRAO and NAOJ.

\end{acknowledgments}

\vspace{5mm}

\facilities{ALMA, MeerKAT}

\software{Astropy \citep{AstropyCollaboration_Astropy_2018, AstropyCollaboration_Astropy_2022},  
          Spectral-cube \citep{Ginsburg_Radio_2015}, 
          Matplotlib (\url{https://matplotlib.org}), 
          lmfit (\url{https://lmfit.github.io/lmfit-py/}), emcee \citep{Foreman-Mackey_emcee_2013}
          }

\appendix

\section{Results of spectral decomposition}

\restartappendixnumbering
{\startlongtable
\begin{deluxetable*}{cccccc}
\tablecaption{Results of spectral decomposition with one or multiple Gaussian components. 
\label{tab:fit}}
\tablehead{ 
\colhead{Region} & 
\colhead{Species} & 
\colhead{Transition} & 
\colhead{$T_{\rm peak}$ (K)} & 
\colhead{$V_{\rm LSR}$ (\kms)} & 
\colhead{FWHM (\kms)}
}
\startdata 
1 & \emethanol & $6_{-1}$--$5_{-1}$ & $0.329\pm0.010$ & $-3.80\pm0.21$ & $8.4\pm1.1$ \\ 
 &  & $6_{0}$--$5_{0}$ & $0.0823\pm0.0040$ & $-4.72\pm0.24$ & $8.63\pm0.62$ \\ 
 &  & $3_{0}$--$2_{-1}$ & $0.0615\pm0.0039$ & $-2.99\pm0.96$ & $12.4\pm1.8$ \\ 
 &  & $6_{2}$--$5_{2}$ & $0.0576\pm0.0040$ & $-3.73\pm0.52$ & $7.5\pm1.1$ \\ 
 &  & $6_{1}$--$5_{1}$ & $0.0295\pm0.0034$ & $-3.59\pm0.83$ & $11.6\pm2.2$ \\ 
 & \amethanol & $6_0$--$5_0$ & $0.416\pm0.009$ & $-3.89\pm0.16$ & $8.87\pm0.64$ \\ 
 & \phhco & $4_{0,4}$--$3_{0,3}$ & $0.174\pm0.004$ & $-2.94\pm0.36$ & $12.42\pm0.65$ \\ 
 &  & $4_{2,2}$--$3_{2,1}$ & $0.0750\pm0.0035$ & $-4.71\pm0.40$ & $10.17\pm0.85$ \\ 
 &  & $4_{2,3}$--$3_{2,2}$ & $0.0759\pm0.0037$ & $-3.85\pm0.50$ & $10.0\pm1.1$ \\ 
 & SO & $7_8$--$6_7$ & $0.153\pm0.004$ & $-4.33\pm0.22$ & $8.95\pm0.51$ \\ 
 & SiO & 7--6 & $0.0615\pm0.0044$ & $-4.00\pm0.36$ & $7.20\pm0.75$ \\ 
\hline
2 & \emethanol & $6_{-1}$--$5_{-1}$ & $0.298\pm0.005$ & $6.77\pm0.08$ & $9.28\pm0.19$ \\ 
 &  & $6_{0}$--$5_{0}$ & $0.0432\pm0.0031$ & $4.78\pm0.48$ & $13.6\pm1.1$ \\ 
 &  & $3_{0}$--$2_{-1}$ & $0.0364\pm0.0034$ & $6.16\pm0.52$ & $11.3\pm1.2$ \\ 
 &  & $6_{2}$--$5_{2}$ & $0.0365\pm0.0036$ & $5.80\pm0.39$ & $7.90\pm0.91$ \\ 
 &  & $6_{1}$--$5_{1}$ & $0.0366\pm0.0049$ & $4.71\pm0.30$ & $4.64\pm0.72^{a}$ \\ 
 & \amethanol & $6_0$--$5_0$ & $0.393\pm0.005$ & $6.69\pm0.06$ & $9.46\pm0.14$ \\ 
 & \phhco & $4_{0,4}$--$3_{0,3}$ & $0.182\pm0.004$ & $6.78\pm0.12$ & $9.78\pm0.27$ \\ 
 &  & $4_{2,2}$--$3_{2,1}$ & $0.0570\pm0.0033$ & $6.05\pm0.33$ & $11.64\pm0.78$ \\ 
 &  & $4_{2,3}$--$3_{2,2}$ & $0.0639\pm0.0034$ & $6.18\pm0.28$ & $10.81\pm0.66$ \\ 
 & SO & $7_8$--$6_7$ & $0.543\pm0.006$ & $5.57\pm0.06$ & $10.48\pm0.13$ \\ 
 & SiO & 7--6 & $0.594\pm0.006$ & $4.90\pm0.05$ & $9.92\pm0.13$ \\ 
\hline
3 & \emethanol & $6_{-1}$--$5_{-1}$ & $0.80\pm0.16$ & $6.10\pm0.40$ & $11.13\pm0.65$ \\ 
low &  & $6_{0}$--$5_{0}$ & $0.114\pm0.014$ & $4.77\pm0.52$ & $12.03\pm0.57$ \\ 
$V_{\rm LSR}$ &  & $3_{0}$--$2_{-1}$ & $0.183\pm0.066$ & $6.76\pm0.47$ & $9.89\pm0.77$ \\ 
component &  & $6_{2}$--$5_{2}$ & $0.132\pm0.017$ & $6.39\pm0.30$ & $11.28\pm0.59$ \\ 
 &  & $6_{1}$--$5_{1}$ & --- & --- & --- \\ 
 & \amethanol & $6_0$--$5_0$ & $0.97\pm0.15$ & $6.12\pm0.32$ & $11.35\pm0.55$ \\ 
 & \phhco & $4_{0,4}$--$3_{0,3}$ & $0.552\pm0.072$ & $5.82\pm0.29$ & $10.06\pm0.22$ \\ 
 &  & $4_{2,2}$--$3_{2,1}$ & $0.150\pm0.021$ & $5.19\pm0.46$ & $11.38\pm0.47$ \\ 
 &  & $4_{2,3}$--$3_{2,2}$ & $0.246\pm0.022$ & $6.28\pm0.18$ & $10.46\pm0.32$ \\ 
 & SO & $7_8$--$6_7$ & $0.380\pm0.016$ & $4.99\pm0.15$ & $12.52\pm0.22$ \\ 
 & SiO & 7--6 & $0.135\pm0.009$ & $4.63\pm0.28$ & $15.74\pm0.57$ \\ 
\hline
3 & \emethanol & $6_{-1}$--$5_{-1}$ & $1.50\pm0.16$ & $7.97\pm0.10$ & $5.83\pm0.34$ \\ 
high &  & $6_{0}$--$5_{0}$ & $0.370\pm0.015$ & $8.05\pm0.06$ & $5.57\pm0.19$ \\ 
$V_{\rm LSR}$ &  & $3_{0}$--$2_{-1}$ & $0.225\pm0.066$ & $8.19\pm0.19$ & $6.07\pm0.66$ \\ 
component &  & $6_{2}$--$5_{2}$ & $0.246\pm0.017$ & $8.19\pm0.08$ & $4.93\pm0.26$ \\ 
 &  & $6_{1}$--$5_{1}$ & $0.182\pm0.004$ & $7.71\pm0.08$ & $6.87\pm0.19$ \\ 
 & \amethanol & $6_0$--$5_0$ & $1.96\pm0.15$ & $7.96\pm0.07$ & $5.87\pm0.26$ \\ 
 & \phhco & $4_{0,4}$--$3_{0,3}$ & $1.07\pm0.08$ & $8.07\pm0.06$ & $6.50\pm0.17$ \\ 
 &  & $4_{2,2}$--$3_{2,1}$ & $0.445\pm0.022$ & $7.98\pm0.06$ & $6.06\pm0.19$ \\ 
 &  & $4_{2,3}$--$3_{2,2}$ & $0.370\pm0.021$ & $8.07\pm0.06$ & $5.03\pm0.19$ \\ 
 & SO & $7_8$--$6_7$ & $0.659\pm0.017$ & $7.62\pm0.04$ & $5.51\pm0.12$ \\ 
 & SiO & 7--6 & $0.109\pm0.010$ & $7.37\pm0.16$ & $5.54\pm0.50$ \\ 
\hline
4 & \emethanol & $6_{-1}$--$5_{-1}$ & $0.661\pm0.091$ & $4.58\pm0.27$ & $11.18\pm0.49$ \\ 
low &  & $6_{0}$--$5_{0}$ & $0.148\pm0.019$ & $4.41\pm0.26$ & $11.70\pm0.62$ \\ 
$V_{\rm LSR}$ &  & $3_{0}$--$2_{-1}$ & $0.132\pm0.024$ & $4.08\pm0.56$ & $10.97\pm0.61$ \\ 
component &  & $6_{2}$--$5_{2}$ & $0.112\pm0.020$ & $5.06\pm0.28$ & $11.51\pm0.90$ \\ 
 &  & $6_{1}$--$5_{1}$ & $0.107\pm0.005$ & $4.99\pm0.17$ & $7.63\pm0.41^a$ \\ 
 & \amethanol & $6_0$--$5_0$ & $0.867\pm0.094$ & $4.58\pm0.22$ & $10.76\pm0.35$ \\ 
 & \phhco & $4_{0,4}$--$3_{0,3}$ & $0.415\pm0.070$ & $3.56\pm0.59$ & $10.77\pm0.35$ \\ 
 &  & $4_{2,2}$--$3_{2,1}$ & $0.171\pm0.031$ & $3.82\pm0.52$ & $11.09\pm0.55$ \\ 
 &  & $4_{2,3}$--$3_{2,2}$ & $0.223\pm0.026$ & $4.71\pm0.22$ & $10.97\pm0.48$ \\ 
 & SO & $7_8$--$6_7$ & $0.263\pm0.022$ & $4.04\pm0.21$ & $12.01\pm0.39$ \\ 
 & SiO & 7--6 & $0.388\pm0.006^{\rm b}$ & $5.41\pm0.08^{\rm b}$ & $10.08\pm0.19^{\rm b}$ \\ 
\hline
4 & \emethanol & $6_{-1}$--$5_{-1}$ & $0.514\pm0.090$ & $6.35\pm0.16$ & $5.66\pm0.59$ \\ 
high &  & $6_{0}$--$5_{0}$ & $0.117\pm0.018$ & $5.96\pm0.16$ & $5.00\pm0.60$ \\ 
$V_{\rm LSR}$ &  & $3_{0}$--$2_{-1}$ & $0.120\pm0.026$ & $6.70\pm0.24$ & $5.71\pm0.77$ \\ 
component &  & $6_{2}$--$5_{2}$ & $0.088\pm0.019$ & $5.98\pm0.19$ & $4.59\pm0.77$ \\ 
 &  & $6_{1}$--$5_{1}$ & --- & --- & --- \\ 
 & \amethanol & $6_0$--$5_0$ & $0.645\pm0.093$ & $6.47\pm0.14$ & $5.57\pm0.47$ \\ 
 & \phhco & $4_{0,4}$--$3_{0,3}$ & $0.566\pm0.083$ & $6.66\pm0.11$ & $6.97\pm0.39$ \\ 
 &  & $4_{2,2}$--$3_{2,1}$ & $0.171\pm0.033$ & $6.35\pm0.20$ & $6.06\pm0.66$ \\ 
 &  & $4_{2,3}$--$3_{2,2}$ & $0.123\pm0.025$ & $6.25\pm0.20$ & $4.99\pm0.71$ \\ 
 & SO & $7_8$--$6_7$ & $0.352\pm0.022$ & $6.01\pm0.07$ & $5.44\pm0.25$ \\ 
 & SiO & 7--6 & --- & --- & --- \\ 
\hline
5 & \emethanol & $6_{-1}$--$5_{-1}$ & $0.302\pm0.014$ & $5.50\pm0.33$ & $15.57\pm0.40$ \\ 
low &  & $6_{0}$--$5_{0}$ & $0.0483\pm0.0094$ & $3.50\pm0.74$ & $11.3\pm3.6$ \\ 
$V_{\rm LSR}$ &  & $3_{0}$--$2_{-1}$ & $0.0695\pm0.0048$ & $6.36\pm0.46$ & $15.03\pm0.75$ \\ 
component &  & $6_{2}$--$5_{2}$ & $0.0774\pm0.0031$ & $7.50\pm0.05$ & $12.54\pm0.54$ \\ 
 &  & $6_{1}$--$5_{1}$ & $0.0389\pm0.0031$ & $7.50\pm0.24$ & $15.0\pm1.2$ \\ 
 & \amethanol & $6_0$--$5_0$ & $0.403\pm0.013$ & $5.75\pm0.24$ & $15.29\pm0.30$ \\ 
 & \phhco & $4_{0,4}$--$3_{0,3}$ & $0.130\pm0.016$ & $4.6\pm1.01$ & $14.56\pm0.98$ \\ 
 &  & $4_{2,2}$--$3_{2,1}$ & $0.070\pm0.012$ & $6.0\pm1.2$ & $14.1\pm1.1$ \\ 
 &  & $4_{2,3}$--$3_{2,2}$ & $0.0725\pm0.0039$ & $6.43\pm0.43$ & $15.86\pm0.70$ \\ 
 & SO & $7_8$--$6_7$ & $0.205\pm0.008$ & $6.05\pm0.35$ & $13.33\pm0.39$ \\ 
 & SiO & 7--6 & $0.392\pm0.010$ & $8.85\pm0.15$ & $12.57\pm0.17$ \\ 
\hline
5 & \emethanol & $6_{-1}$--$5_{-1}$ & $0.220\pm0.018$ & $10.40\pm0.16$ & $6.13\pm0.50$ \\ 
high &  & $6_{0}$--$5_{0}$ & $0.0840\pm0.026$ & $10.63\pm0.51$ & $6.9\pm1.0$ \\ 
$V_{\rm LSR}$ &  & $3_{0}$--$2_{-1}$ & $0.0371\pm0.007$ & $10.46\pm0.33$ & $4.30\pm0.96$ \\ 
component &  & $6_{2}$--$5_{2}$ & --- & --- & --- \\ 
 &  & $6_{1}$--$5_{1}$ & --- & --- & --- \\ 
 & \amethanol & $6_0$--$5_0$ & $0.262\pm0.018$ & $10.55\pm0.13$ & $5.94\pm0.40$ \\ 
 & \phhco & $4_{0,4}$--$3_{0,3}$ & $0.198\pm0.026$ & $10.12\pm0.15$ & $7.57\pm0.60$ \\ 
 &  & $4_{2,2}$--$3_{2,1}$ & $0.054\pm0.016$ & $10.85\pm0.40$ & $6.9\pm1.4$ \\ 
 &  & $4_{2,3}$--$3_{2,2}$ & $0.0449\pm0.006$ & $11.23\pm0.25$ & $4.10\pm0.72$ \\ 
 & SO & $7_8$--$6_7$ & $0.286\pm0.014$ & $11.32\pm0.06$ & $5.85\pm0.23$ \\ 
 & SiO & 7--6 & $0.147\pm0.013$ & $12.72\pm0.13$ & $5.13\pm0.43$ \\ 
\hline
6 & \emethanol & $6_{-1}$--$5_{-1}$ & $0.198\pm0.004$ & $7.79\pm0.21$ & $19.19\pm0.55$ \\ 
&  & $6_{0}$--$5_{0}$ & $0.0520\pm0.003$ & $8.07\pm0.51$ & $16.8\pm1.2$ \\ 
 &  & $3_{0}$--$2_{-1}$ & $0.0613\pm0.004$ & $8.08\pm0.44$ & $14.9\pm1.0$ \\ 
 &  & $6_{2}$--$5_{2}$ & $0.0441\pm0.003$ & $7.32\pm0.50$ & $15.9\pm1.2$ \\ 
 &  & $6_{1}$--$5_{1}$ & $0.0316\pm0.004$ & $8.10\pm0.67$ & $12.3\pm1.6$ \\ 
 & \amethanol & $6_0$--$5_0$ & $0.262\pm0.004$ & $6.89\pm0.21$ & $17.14\pm0.52$ \\ 
 & \phhco & $4_{0,4}$--$3_{0,3}$ & $0.122\pm0.003$ & $6.45\pm0.23$ & $19.00\pm0.60$ \\ 
 &  & $4_{2,2}$--$3_{2,1}$ & $0.0574\pm0.003$ & $7.27\pm0.41$ & $17.24\pm0.96$ \\ 
 &  & $4_{2,3}$--$3_{2,2}$ & $0.0469\pm0.003$ & $7.11\pm0.56$ & $20.7\pm1.3$ \\ 
 & SO & $7_8$--$6_7$ & $0.0740\pm0.003$ & $8.67\pm0.36$ & $21.85\pm0.87$ \\ 
 & SiO & 7--6 & $0.133\pm0.003$ & $12.01\pm0.21$ & $16.51\pm0.49$ \\ 
\hline
\enddata
\tablecomments{
$^{\rm a}$ Not used for the non-LTE analysis because its FWHM is much small than those of the other lines. 
$^{\rm b}$ The \vlsr\ of this SiO emission is slightly different from the other lines in this component. Considering its large FWHM, we classify it into the low \vlsr\ component. There may still be some contribution from the high \vlsr\ component which is not distinguishable. 
}
\end{deluxetable*}
}

\section{Supplementary figures}

\restartappendixnumbering

\begin{figure*}[htbp]
\centering
\includegraphics[width=0.95\textwidth]{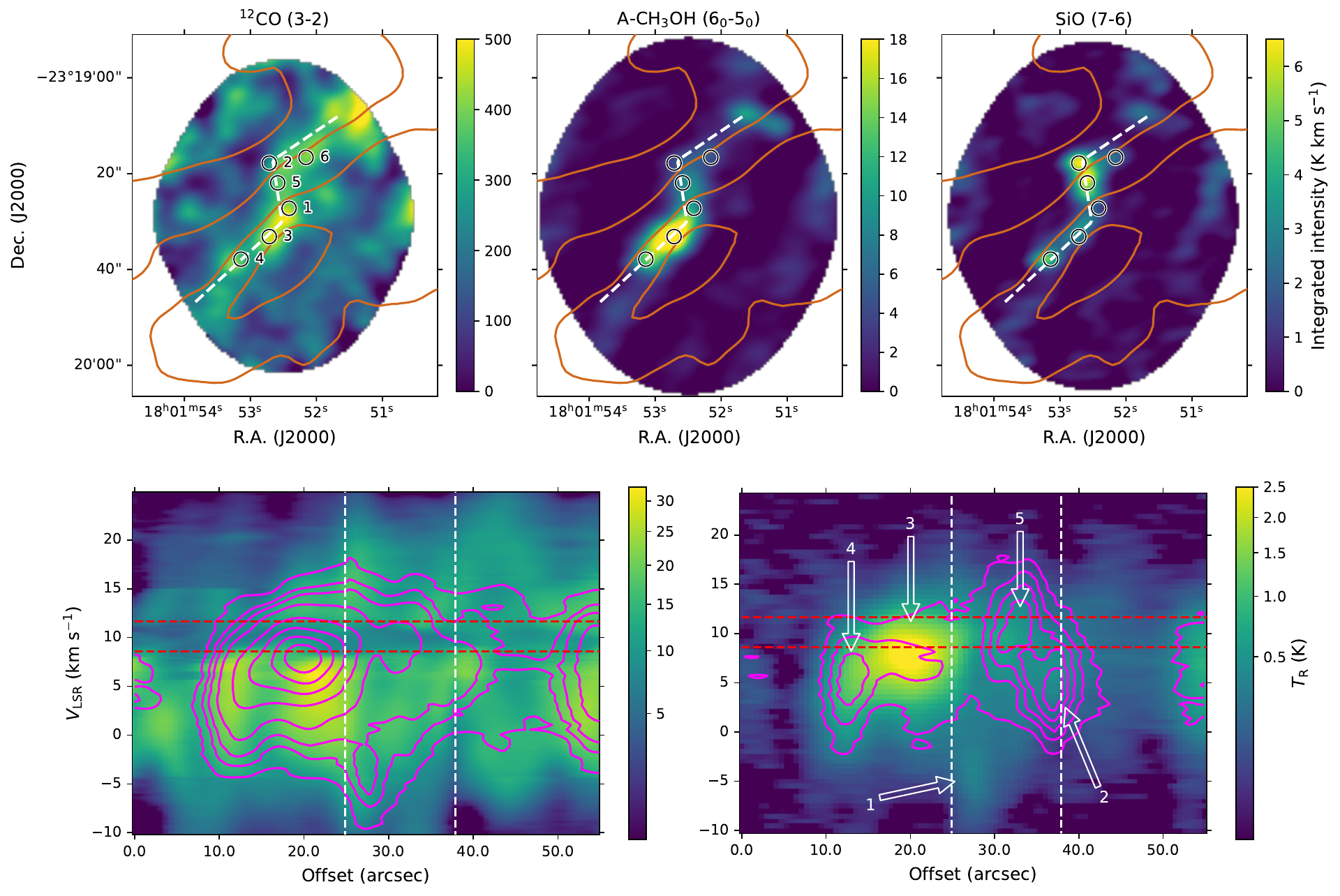}
\caption{ (Upper panels) Integrated intensity maps of the \coa\ (3--2), \amethanol\ ($6_0$--$5_0$), and SiO (7--6) lines in velocity range $-12$ to $+20$ \kms, overlaid with orange contours of MeerKAT 1.3 GHz radio continuum emission in levels of 4, 7, 10 and 13 mJy $\rm beam^{-1}$.
The black circles are regions 1--6. 
The dashed white line in each panel show the path along which we extract the position-velocity diagrams in the lower panels. 
(Lower panels)
Position-velocity diagrams of \coa\ (3--2) overlaid with contours of \amethanol\ ($6_0$--$5_0$) (lower left panel, in levels of 0.1, 0.2, 0.3, 0.5, 1, 1.5, 2, 2.5 K) and \amethanol\ ($6_0$--$5_0$) overlaid with contours of SiO (7--6) (lower right panel, in levels of 0.1, 0.2, 0.3, 0.4, 0.5 K) along the dashed white line shown in the upper panels starting from southeast to northwest with a width of $3^{\prime\prime}$. 
The horizontal dashed red lines mark the range of the \vlsr\ of the OH masers (+8.58 to +11.66 \kms), while the vertical dashed white lines mark the two turning points of the path along which we subtract the PV diagram. 
The approximate positions of regions 1--5 are marked by arrows in the lower right panel. 
\label{fig:pv}
}
\end{figure*}

\bibliography{2026_article_W28F_ALMA}{}

@article{Anderl_APEX_2014,
  title = {{{APEX}} Observations of Supernova Remnants. {{I}}. {{Non-stationary}} Magnetohydrodynamic Shocks in {{W44}}},
  author = {Anderl, S. and Gusdorf, A. and G{\"u}sten, R.},
  year = 2014,
  journal = {A\&A},
  volume = {569},
  pages = {A81},
  doi = {10.1051/0004-6361/201423561},
  langid = {english}
}

@article{Arikawa_Shocked_1999,
  title = {Shocked {{Molecular Gas Associated}} with the {{Supernova Remnant W28}}},
  author = {Arikawa, Yuji and Tatematsu, Ken'ichi and Sekimoto, Yutaro and Takahashi, Tadayuki},
  year = 1999,
  journal = {PASJ},
  volume = {51},
  number = {4},
  pages = {L7-L10},
  doi = {10.1093/pasj/51.4.L7},
  langid = {english}
}

@article{Armijos-Abendano_Structure_2020,
  title = {Structure and Kinematics of Shocked Gas in {{Sgr B2}}: Further Evidence of a Cloud-Cloud Collision from {{SiO}} Emission Maps},
  shorttitle = {Structure and Kinematics of Shocked Gas in {{Sgr B2}}},
  author = {{Armijos-Abenda{\~n}o}, J. and {Banda-Barrag{\'a}n}, W. E. and {Mart{\'i}n-Pintado}, J. and D{\'e}nes, H. and Federrath, C. and {Requena-Torres}, M. A.},
  year = 2020,
  journal = {MNRAS},
  volume = {499},
  pages = {4918--4939},
  publisher = {OUP},
  doi = {10.1093/mnras/staa3119}
}

@article{AstropyCollaboration_Astropy_2018,
  title = {The {{Astropy Project}}: {{Building}} an {{Open-science Project}} and {{Status}} of the v2.0 {{Core Package}}},
  shorttitle = {The {{Astropy Project}}},
  author = {{Astropy Collaboration} and {Price-Whelan}, A. M. and Sip{\H o}cz, B. M. and G{\"u}nther, H. M. and Lim, P. L. and Crawford, S. M. and Conseil, S. and Shupe, D. L. and Craig, M. W. and Dencheva, N. and Ginsburg, A. and VanderPlas, J. T. and Bradley, L. D. and {P{\'e}rez-Su{\'a}rez}, D. and {de Val-Borro}, M. and Aldcroft, T. L. and Cruz, K. L. and Robitaille, T. P. and Tollerud, E. J. and Ardelean, C. and Babej, T. and Bach, Y. P. and Bachetti, M. and Bakanov, A. V. and Bamford, S. P. and Barentsen, G. and Barmby, P. and Baumbach, A. and Berry, K. L. and Biscani, F. and Boquien, M. and Bostroem, K. A. and Bouma, L. G. and Brammer, G. B. and Bray, E. M. and Breytenbach, H. and Buddelmeijer, H. and Burke, D. J. and Calderone, G. and Cano Rodr{\'i}guez, J. L. and Cara, M. and Cardoso, J. V. M. and Cheedella, S. and Copin, Y. and Corrales, L. and Crichton, D. and D'Avella, D. and Deil, C. and Depagne, {\'E}. and Dietrich, J. P. and Donath, A. and Droettboom, M. and Earl, N. and Erben, T. and Fabbro, S. and Ferreira, L. A. and Finethy, T. and Fox, R. T. and Garrison, L. H. and Gibbons, S. L. J. and Goldstein, D. A. and Gommers, R. and Greco, J. P. and Greenfield, P. and Groener, A. M. and Grollier, F. and Hagen, A. and Hirst, P. and Homeier, D. and Horton, A. J. and Hosseinzadeh, G. and Hu, L. and Hunkeler, J. S. and Ivezi{\'c}, {\v Z}. and Jain, A. and Jenness, T. and Kanarek, G. and Kendrew, S. and Kern, N. S. and Kerzendorf, W. E. and Khvalko, A. and King, J. and Kirkby, D. and Kulkarni, A. M. and Kumar, A. and Lee, A. and Lenz, D. and Littlefair, S. P. and Ma, Z. and Macleod, D. M. and Mastropietro, M. and McCully, C. and Montagnac, S. and Morris, B. M. and Mueller, M. and Mumford, S. J. and Muna, D. and Murphy, N. A. and Nelson, S. and Nguyen, G. H. and Ninan, J. P. and N{\"o}the, M. and Ogaz, S. and Oh, S. and Parejko, J. K. and Parley, N. and Pascual, S. and Patil, R. and Patil, A. A. and Plunkett, A. L. and Prochaska, J. X. and Rastogi, T. and Reddy Janga, V. and Sabater, J. and Sakurikar, P. and Seifert, M. and Sherbert, L. E. and {Sherwood-Taylor}, H. and Shih, A. Y. and Sick, J. and Silbiger, M. T. and Singanamalla, S. and Singer, L. P. and Sladen, P. H. and Sooley, K. A. and Sornarajah, S. and Streicher, O. and Teuben, P. and Thomas, S. W. and Tremblay, G. R. and Turner, J. E. H. and Terr{\'o}n, V. and {van Kerkwijk}, M. H. and {de la Vega}, A. and Watkins, L. L. and Weaver, B. A. and Whitmore, J. B. and Woillez, J. and Zabalza, V. and {Astropy Contributors}},
  year = 2018,
  journal = {ApJ},
  volume = {156},
  pages = {123},
  doi = {10.3847/1538-3881/aabc4f}
}

@article{AstropyCollaboration_Astropy_2022,
  title = {The {{Astropy Project}}: {{Sustaining}} and {{Growing}} a {{Community-oriented Open-source Project}} and the {{Latest Major Release}} (v5.0) of the {{Core Package}}},
  shorttitle = {The {{Astropy Project}}},
  author = {{Astropy Collaboration} and {Price-Whelan}, Adrian M. and Lim, Pey Lian and Earl, Nicholas and Starkman, Nathaniel and Bradley, Larry and Shupe, David L. and Patil, Aarya A. and Corrales, Lia and Brasseur, C. E. and N{\"o}the, Maximilian and Donath, Axel and Tollerud, Erik and Morris, Brett M. and Ginsburg, Adam and Vaher, Eero and Weaver, Benjamin A. and Tocknell, James and Jamieson, William and {van Kerkwijk}, Marten H. and Robitaille, Thomas P. and Merry, Bruce and Bachetti, Matteo and G{\"u}nther, H. Moritz and Aldcroft, Thomas L. and {Alvarado-Montes}, Jaime A. and Archibald, Anne M. and B{\'o}di, Attila and Bapat, Shreyas and Barentsen, Geert and Baz{\'a}n, Juanjo and Biswas, Manish and Boquien, M{\'e}d{\'e}ric and Burke, D. J. and Cara, Daria and Cara, Mihai and Conroy, Kyle E. and Conseil, Simon and Craig, Matthew W. and Cross, Robert M. and Cruz, Kelle L. and D'Eugenio, Francesco and Dencheva, Nadia and Devillepoix, Hadrien A. R. and Dietrich, J{\"o}rg P. and Eigenbrot, Arthur Davis and Erben, Thomas and Ferreira, Leonardo and {Foreman-Mackey}, Daniel and Fox, Ryan and Freij, Nabil and Garg, Suyog and Geda, Robel and Glattly, Lauren and Gondhalekar, Yash and Gordon, Karl D. and Grant, David and Greenfield, Perry and Groener, Austen M. and Guest, Steve and Gurovich, Sebastian and Handberg, Rasmus and Hart, Akeem and {Hatfield-Dodds}, Zac and Homeier, Derek and Hosseinzadeh, Griffin and Jenness, Tim and Jones, Craig K. and Joseph, Prajwel and Kalmbach, J. Bryce and Karamehmetoglu, Emir and Ka{\l}uszy{\'n}ski, Miko{\l}aj and Kelley, Michael S. P. and Kern, Nicholas and Kerzendorf, Wolfgang E. and Koch, Eric W. and Kulumani, Shankar and Lee, Antony and Ly, Chun and Ma, Zhiyuan and MacBride, Conor and Maljaars, Jakob M. and Muna, Demitri and Murphy, N. A. and Norman, Henrik and O'Steen, Richard and Oman, Kyle A. and Pacifici, Camilla and Pascual, Sergio and {Pascual-Granado}, J. and Patil, Rohit R. and Perren, Gabriel I. and Pickering, Timothy E. and Rastogi, Tanuj and Roulston, Benjamin R. and Ryan, Daniel F. and Rykoff, Eli S. and Sabater, Jose and Sakurikar, Parikshit and Salgado, Jes{\'u}s and Sanghi, Aniket and Saunders, Nicholas and Savchenko, Volodymyr and Schwardt, Ludwig and {Seifert-Eckert}, Michael and Shih, Albert Y. and Jain, Anany Shrey and Shukla, Gyanendra and Sick, Jonathan and Simpson, Chris and Singanamalla, Sudheesh and Singer, Leo P. and Singhal, Jaladh and Sinha, Manodeep and Sip{\H o}cz, Brigitta M. and Spitler, Lee R. and Stansby, David and Streicher, Ole and {\v S}umak, Jani and Swinbank, John D. and Taranu, Dan S. and Tewary, Nikita and Tremblay, Grant R. and {de Val-Borro}, Miguel and Van Kooten, Samuel J. and Vasovi{\'c}, Zlatan and Verma, Shresth and {de Miranda Cardoso}, Jos{\'e} Vin{\'i}cius and Williams, Peter K. G. and Wilson, Tom J. and Winkel, Benjamin and {Wood-Vasey}, W. M. and Xue, Rui and Yoachim, Peter and Zhang, Chen and Zonca, Andrea and {Astropy Project Contributors}},
  year = 2022,
  journal = {ApJ},
  volume = {935},
  pages = {167},
  doi = {10.3847/1538-4357/ac7c74}
}

@article{Bachiller_Methanol_1995,
  title = {Methanol Enhancemant in Young Bipolar Outflows.},
  author = {Bachiller, R. and Liechti, S. and Walmsley, C. M. and Colomer, F.},
  year = 1995,
  journal = {A\&A},
  volume = {295},
  pages = {L51}
}

@article{Balanca_Rotationally_2018,
  title = {Rotationally Inelastic Collisions of {{SiO}} with {{H2}}},
  author = {Balan{\c c}a, Christian and Dayou, Fabrice and Faure, Alexandre and Wiesenfeld, Laurent and Feautrier, Nicole},
  year = 2018,
  journal = {MNRAS},
  volume = {479},
  pages = {2692--2701},
  publisher = {OUP},
  doi = {10.1093/mnras/sty1681}
}

@article{Bergin_Postshock_1998,
  title = {The {{Postshock Chemical Lifetimes}} of {{Outflow Tracers}} and a {{Possible New Mechanism}} to {{Produce Water Ice Mantles}}},
  author = {Bergin, Edwin A. and Melnick, Gary J. and Neufeld, David A.},
  year = 1998,
  journal = {ApJ},
  volume = {499},
  pages = {777--792},
  publisher = {IOP},
  doi = {10.1086/305656}
}

@article{Bisschop_Testing_2007,
  title = {Testing Grain-Surface Chemistry in Massive Hot-Core Regions},
  author = {Bisschop, S. E. and J{\o}rgensen, J. K. and {van Dishoeck}, E. F. and {de Wachter}, E. B. M.},
  year = 2007,
  journal = {A\&A},
  volume = {465},
  pages = {913--929},
  publisher = {EDP},
  doi = {10.1051/0004-6361:20065963}
}

@article{Bouvier_ALCHEMI_2024,
  title = {An {{ALCHEMI}} Inspection of Sulphur-Bearing Species towards the Central Molecular Zone of {{NGC}} 253},
  author = {Bouvier, M. and Viti, S. and Behrens, E. and Butterworth, J. and Huang, K.-Y. and Mangum, J. G. and Harada, N. and Mart{\'i}n, S. and Rivilla, V. M. and Muller, S. and Sakamoto, K. and Yoshimura, Y. and Tanaka, K. and Nakanishi, K. and {Herrero-Illana}, R. and Colzi, L. and Gorski, M. D. and Henkel, C. and Humire, P. K. and Meier, D. S. and {van der Werf}, P. P. and Yan, Y. T.},
  year = 2024,
  journal = {A\&A},
  volume = {689},
  pages = {A64},
  publisher = {EDP},
  doi = {10.1051/0004-6361/202449186}
}

@article{Burkhardt_Modeling_2019,
  title = {Modeling {{C-shock Chemistry}} in {{Isolated Molecular Outflows}}},
  author = {Burkhardt, Andrew M. and Shingledecker, Christopher N. and Le Gal, Romane and McGuire, Brett A. and Remijan, Anthony J. and Herbst, Eric},
  year = 2019,
  journal = {ApJ},
  volume = {881},
  pages = {32},
  doi = {10.3847/1538-4357/ab2be8}
}

@article{CASATeam_CASA_2022,
  title = {{{CASA}}, the {{Common Astronomy Software Applications}} for {{Radio Astronomy}}},
  author = {{CASA Team} and Bean, Ben and Bhatnagar, Sanjay and Castro, Sandra and Donovan Meyer, Jennifer and Emonts, Bjorn and Garcia, Enrique and Garwood, Robert and Golap, Kumar and Gonzalez Villalba, Justo and Harris, Pamela and Hayashi, Yohei and Hoskins, Josh and Hsieh, Mingyu and Jagannathan, Preshanth and Kawasaki, Wataru and Keimpema, Aard and Kettenis, Mark and Lopez, Jorge and Marvil, Joshua and Masters, Joseph and McNichols, Andrew and Mehringer, David and Miel, Renaud and Moellenbrock, George and Montesino, Federico and Nakazato, Takeshi and Ott, Juergen and Petry, Dirk and Pokorny, Martin and Raba, Ryan and Rau, Urvashi and Schiebel, Darrell and Schweighart, Neal and Sekhar, Srikrishna and Shimada, Kazuhiko and Small, Des and Steeb, Jan-Willem and Sugimoto, Kanako and Suoranta, Ville and Tsutsumi, Takahiro and {van Bemmel}, Ilse M. and Verkouter, Marjolein and Wells, Akeem and Xiong, Wei and Szomoru, Arpad and Griffith, Morgan and Glendenning, Brian and Kern, Jeff},
  year = 2022,
  journal = {PASP},
  volume = {134},
  pages = {114501},
  publisher = {IOP},
  doi = {10.1088/1538-3873/ac9642}
}

@article{Caselli_Grain-grain_1997,
  title = {Grain-Grain Collisions and Sputtering in Oblique {{C-type}} Shocks.},
  author = {Caselli, P. and Hartquist, T. W. and Havnes, O.},
  year = 1997,
  journal = {A\&A},
  volume = {322},
  pages = {296--301}
}

@article{Ceccarelli_Supernova-enhanced_2011,
  title = {Supernova-Enhanced {{Cosmic-Ray Ionization}} and {{Induced Chemistry}} in a {{Molecular Cloud}} of {{W51C}}},
  author = {Ceccarelli, C. and {Hily-Blant}, P. and Montmerle, T. and Dubus, G. and Gallant, Y. and Fiasson, A.},
  year = 2011,
  journal = {ApJL},
  volume = {740},
  pages = {L4},
  doi = {10.1088/2041-8205/740/1/L4}
}

@article{Chi_complete_2026,
  title = {A Complete {{X-ray}} View of Supernova Remnant {{W28}} with the {{Einstein Probe}}: {{Spatial}} Distribution of Parameters, and the Origin of the Thermal-Composite Morphology},
  shorttitle = {A Complete {{X-ray}} View of Supernova Remnant {{W28}} with the {{Einstein Probe}}},
  author = {Chi, Yi-Heng and Zhou, Ping and Chen, Yang and Sun, Lei and Li, Chengkui and Jia, Shumei and Chen, Yong and Ge, Chong and Yuan, Weimin},
  year = 2026,
  journal = {A\&A},
  volume = {709},
  pages = {A257},
  publisher = {EDP},
  doi = {10.1051/0004-6361/202556500}
}

@article{Claussen_Polarization_1997,
  title = {Polarization {{Observations}} of 1720 {{MHz OH Masers}} toward the {{Three Supernova Remnants W28}}, {{W44}}, and {{IC}} 443},
  author = {Claussen, M. J. and Frail, D. A. and Goss, W. M. and Gaume, R. A.},
  year = 1997,
  journal = {ApJ},
  volume = {489},
  pages = {143--159},
  doi = {10.1086/304784}
}

@article{Codella_SOLIS_2025a,
  title = {{{SOLIS}}: {{XIX}}. {{The}} Chemically Rich {{SVS13-B}} Protostellar Jet},
  shorttitle = {{{SOLIS}}},
  author = {Codella, C. and Bianchi, E. and Podio, L. and De Simone, M. and {L{\'o}pez-Sepulcre}, A. and Ceccarelli, C. and Caselli, P.},
  year = 2025,
  journal = {A\&A},
  volume = {696},
  pages = {A219},
  publisher = {EDP},
  doi = {10.1051/0004-6361/202453621}
}

@article{Cosentino_Interstellar_2019,
  title = {Interstellar {{Plunging Waves}}: {{ALMA Resolves}} the {{Physical Structure}} of {{Nonstationary MHD Shocks}}},
  shorttitle = {Interstellar {{Plunging Waves}}},
  author = {Cosentino, Giuliana and {Jim{\'e}nez-Serra}, Izaskun and Caselli, Paola and Henshaw, Jonathan D. and Barnes, Ashley T. and Tan, Jonathan C. and Viti, Serena and Fontani, Francesco and Wu, Benjamin},
  year = 2019,
  journal = {ApJ},
  volume = {881},
  pages = {L42},
  doi = {10.3847/2041-8213/ab38c5}
}

@article{Cosentino_Negative_2022,
  title = {Negative and {{Positive Feedback}} from a {{Supernova Remnant}} with {{SHREC}}: {{A}} Detailed {{Study}} of the {{Shocked Gas}} in {{IC443}}},
  shorttitle = {Negative and {{Positive Feedback}} from a {{Supernova Remnant}} with {{SHREC}}},
  author = {Cosentino, G. and {Jim{\'e}nez-Serra}, I. and Tan, J. C. and Henshaw, J. D. and Barnes, A. T. and Law, C.-Y. and Zeng, S. and Fontani, F. and Caselli, P. and Viti, S. and Zahorecz, S. and {Rico-Villas}, F. and Meg{\'i}as, A. and Miceli, M. and Orlando, S. and Ustamujic, S. and Greco, E. and Peres, G. and Bocchino, F. and Fedriani, R. and Gorai, P. and Testi, L. and {Mart{\'i}n-Pintado}, J.},
  year = 2022,
  journal = {MNRAS},
  volume = {511},
  number = {1},
  eprint = {2201.03008},
  pages = {953--963},
  doi = {10.1093/mnras/stac070},
  archiveprefix = {arXiv},
  langid = {english}
}

@article{Cosentino_Widespread_2018,
  title = {Widespread {{SiO}} and {{CH3OH}} Emission in Filamentary Infrared Dark Clouds},
  author = {Cosentino, G. and {Jim{\'e}nez-Serra}, I. and Henshaw, J. D. and Caselli, P. and Viti, S. and Barnes, A. T. and Fontani, F. and Tan, J. C. and Pon, A.},
  year = 2018,
  journal = {MNRAS},
  volume = {474},
  pages = {3760--3781},
  publisher = {OUP},
  doi = {10.1093/mnras/stx3013}
}

@article{Cuppen_Microscopic_2009,
  title = {Microscopic Simulation of Methanol and Formaldehyde Ice Formation in Cold Dense Cores},
  author = {Cuppen, H. M. and {van Dishoeck}, E. F. and Herbst, E. and Tielens, A. G. G. M.},
  year = 2009,
  journal = {A\&A},
  volume = {508},
  pages = {275--287},
  publisher = {EDP},
  doi = {10.1051/0004-6361/200913119}
}

@article{Dagdigian_Rotational_2024,
  title = {Rotational Excitation of Methanol in Collisions with Molecular Hydrogen},
  author = {Dagdigian, Paul J.},
  year = 2024,
  journal = {MNRAS},
  volume = {527},
  pages = {2209--2213},
  publisher = {OUP},
  doi = {10.1093/mnras/stad3303}
}

@article{DellOva_Interstellar_2020,
  title = {Interstellar Anatomy of the {{TeV}} Gamma-Ray Peak in the {{IC443}} Supernova Remnant},
  author = {Dell'Ova, P. and Gusdorf, A. and Gerin, M. and Riquelme, D. and G{\"u}sten, R. and {Noriega-Crespo}, A. and Tram, L. N. and Houde, M. and Guillard, P. and Lehmann, A. and Lesaffre, P. and Louvet, F. and Marcowith, A. and Padovani, M.},
  year = 2020,
  journal = {A\&A},
  volume = {644},
  pages = {A64},
  doi = {10.1051/0004-6361/202038339},
  langid = {english}
}

@article{Draine_Theory_1993,
  title = {Theory of Interstellar Shocks},
  author = {Draine, Bruce T. and McKee, Christopher F.},
  year = 1993,
  journal = {ARA\&A},
  volume = {31},
  pages = {373--432},
  doi = {10.1146/annurev.aa.31.090193.002105},
  langid = {english}
}

@article{Flower_Excitation_2010,
  title = {Excitation and Emission of {{H2}}, {{CO}} and {{H2O}} Molecules in Interstellar Shock Waves},
  author = {Flower, D. R. and Pineau Des For{\^e}ts, G.},
  year = 2010,
  journal = {Monthly Notices of the Royal Astronomical Society},
  volume = {406},
  pages = {1745--1758},
  publisher = {OUP},
  doi = {10.1111/j.1365-2966.2010.16834.x}
}

@article{Flower_Theoretical_1985,
  title = {Theoretical Studies of Interstellar Molecular Shocks. {{I}} - {{General}} Formulation and Effects of the Ion-Molecule Chemistry},
  author = {Flower, D. R. and {Pineau des For{\^e}ts}, G. and Hartquist, T. W.},
  year = 1985,
  journal = {MNRAS},
  volume = {216},
  pages = {775--794},
  doi = {10.1093/mnras/216.4.775}
}

@article{Foreman-Mackey_emcee_2013,
  title = {Emcee: {{The MCMC Hammer}}},
  shorttitle = {Emcee},
  author = {{Foreman-Mackey}, Daniel and Hogg, David W. and Lang, Dustin and Goodman, Jonathan},
  year = 2013,
  journal = {PASP},
  volume = {125},
  pages = {306},
  publisher = {IOP},
  doi = {10.1086/670067}
}

@article{Fuchs_Hydrogenation_2009,
  title = {Hydrogenation Reactions in Interstellar {{CO}} Ice Analogues. {{A}} Combined Experimental/Theoretical Approach},
  author = {Fuchs, G. W. and Cuppen, H. M. and Ioppolo, S. and Romanzin, C. and Bisschop, S. E. and Andersson, S. and {van Dishoeck}, E. F. and Linnartz, H.},
  year = 2009,
  journal = {A\&A},
  volume = {505},
  pages = {629--639},
  doi = {10.1051/0004-6361/200810784}
}

@article{Fuente_Gas_2023,
  title = {Gas Phase {{Elemental}} Abundances in {{Molecular cloudS}} ({{GEMS}}). {{VII}}. {{Sulfur}} Elemental Abundance},
  author = {Fuente, A. and {Rivi{\`e}re-Marichalar}, P. and {Beitia-Antero}, L. and Caselli, P. and Wakelam, V. and Esplugues, G. and {Rodr{\'i}guez-Baras}, M. and {Navarro-Almaida}, D. and Gerin, M. and Kramer, C. and Bachiller, R. and Goicoechea, J. R. and {Jim{\'e}nez-Serra}, I. and Loison, J. C. and Ivlev, A. and {Mart{\'i}n-Dom{\'e}nech}, R. and Spezzano, S. and Roncero, O. and {Mu{\~n}oz-Caro}, G. and Cazaux, S. and Marcelino, N.},
  year = 2023,
  journal = {A\&A},
  volume = {670},
  pages = {A114},
  doi = {10.1051/0004-6361/202244843},
  langid = {english}
}

@article{Garay_Methanol_2002,
  title = {Methanol and {{Silicon Monoxide Observations}} toward {{Bipolar Outflows Associated}} with {{Class}} 0 {{Objects}}},
  author = {Garay, Guido and Mardones, Diego and Rodr{\'i}guez, Luis F. and Caselli, Paola and Bourke, Tyler L.},
  year = 2002,
  journal = {ApJ},
  volume = {567},
  pages = {980--998},
  publisher = {IOP},
  doi = {10.1086/338668}
}

@article{Garay_Molecular_1998,
  title = {Molecular {{Abundance Enhancements}} in the {{Highly Collimated Bipolar Outflow BHR}} 71},
  author = {Garay, Guido and K{\"o}hnenkamp, Ive and Bourke, Tyler L. and Rodr{\'i}guez, L. F. and Lehtinen, Kimmo K.},
  year = 1998,
  journal = {ApJ},
  volume = {509},
  pages = {768--784},
  publisher = {IOP},
  doi = {10.1086/306534}
}

@article{Garay_Silicon_2000,
  title = {Silicon {{Monoxide}} and {{Methanol Emission}} from the {{NGC}} 2071 {{Molecular Outflow}}},
  author = {Garay, Guido and Mardones, Diego and Rodr{\'i}guez, L. F.},
  year = 2000,
  journal = {ApJ},
  volume = {545},
  pages = {861--873},
  doi = {10.1086/317853}
}

@inproceedings{Ginsburg_Radio_2015,
  title = {Radio {{Astronomy Tools}} in {{Python}}: {{Spectral-cube}}, Pvextractor, and More},
  shorttitle = {Radio {{Astronomy Tools}} in {{Python}}},
  booktitle = {{{ASP Conference Series}}},
  author = {Ginsburg, A. and Robitaille, T. and Beaumont, C. and Rosolowsky, E. and Leroy, A. and Brogan, C. and Hunter, T. and Teuben, P. and Brisbin, D.},
  year = 2015,
  volume = {499},
  pages = {363--364},
  publisher = {Astronomical Society of the Pacific},
  address = {San Francisco}
}

@article{Goedhart_SARAO_2024,
  title = {The {{SARAO MeerKAT}} 1.3 {{GHz Galactic Plane Survey}}},
  author = {Goedhart, S. and Cotton, W. D. and Camilo, F. and Thompson, M. A. and Umana, G. and Bietenholz, M. and Woudt, P. A. and Anderson, L. D. and Bordiu, C. and Buckley, D. A. H. and Buemi, C. S. and Bufano, F. and Cavallaro, F. and Chen, H. and Chibueze, J. O. and Egbo, D. and Frank, B. S. and Hoare, M. G. and Ingallinera, A. and Irabor, T. and {Kraan-Korteweg}, R. C. and Kurapati, S. and Leto, P. and Loru, S. and Mutale, M. and Obonyo, W. O. and Plavin, A. and Rajohnson, S. H. A. and Rigby, A. and Riggi, S. and Seidu, M. and Serra, P. and Smart, B. M. and Stappers, B. W. and Steyn, N. and Surnis, M. and Trigilio, C. and Williams, G. M. and Abbott, T. D. and Adam, R. M. and Asad, K. M. B. and Baloyi, T. and Bauermeister, E. F. and Bennet, T. G. H. and Bester, H. and Botha, A. G. and Brederode, L. R. S. and Buchner, S. and Burger, J. P. and Cheetham, T. and Cloete, K. and {de Villiers}, M. S. and {de Villiers}, D. I. L. and {du Toit}, L. J. and Esterhuyse, S. W. P. and Fanaroff, B. L. and Fourie, D. J. and Gamatham, R. R. G. and Gatsi, T. G. and Geyer, M. and Gouws, M. and Gumede, S. C. and Heywood, I. and Hokwana, A. and Hoosen, S. W. and Horn, D. M. and Horrell, L. M. G. and Hugo, B. V. and Isaacson, A. I. and J{\'o}zsa, G. I. G. and Jonas, J. L. and Jordaan, J. D. B. L. and Joubert, A. F. and Julie, R. P. M. and Kapp, F. B. and Kriek, N. and Kriel, H. and Krishnan, V. K. and Kusel, T. W. and Legodi, L. S. and Lehmensiek, R. and Lord, R. T. and Macfarlane, P. S. and Magnus, L. G. and Magozore, C. and Main, J. P. L. and Malan, J. A. and Manley, J. R. and Marais, S. J. and Maree, M. D. J. and Martens, A. and Maruping, P. and McAlpine, K. and Merry, B. C. and Mgodeli, M. and Millenaar, R. P. and Mokone, O. J. and Monama, T. E. and New, W. S. and Ngcebetsha, B. and Ngoasheng, K. J. and Nicolson, G. D. and Ockards, M. T. and Oozeer, N. and Passmoor, S. S. and Patel, A. A. and {Peens-Hough}, A. and Perkins, S. J. and Ramaila, A. J. T. and Ratcliffe, S. M. and Renil, R. and Richter, L. L. and Salie, S. and Sambu, N. and Schollar, C. T. G. and Schwardt, L. C. and Schwartz, R. L. and Serylak, M. and Siebrits, R. and Sirothia, S. K. and Slabber, M. J. and Smirnov, O. M. and Tiplady, A. J. and {van Balla}, T. J. and {van der Byl}, A. and Van Tonder, V. and Venter, A. J. and Venter, M. and Welz, M. G. and Williams, L. P.},
  year = 2024,
  journal = {MNRAS},
  volume = {531},
  pages = {649--681},
  publisher = {OUP},
  doi = {10.1093/mnras/stae1166}
}

@article{Gomez-Ruiz_1157_2013,
  title = {The {{L}} 1157 Protostellar Outflow Imaged with the {{Submillimeter Array}}},
  author = {{G{\'o}mez-Ruiz}, A. I. and Hirano, N. and Leurini, S. and Liu, S. -Y.},
  year = 2013,
  journal = {A\&A},
  volume = {558},
  pages = {A94},
  doi = {10.1051/0004-6361/201118473}
}

@article{Guillet_Shocks_2011,
  title = {Shocks in Dense Clouds. {{III}}. {{Dust}} Processing and Feedback Effects in {{C-type}} Shocks},
  author = {Guillet, V. and Pineau Des For{\^e}ts, G. and Jones, A. P.},
  year = 2011,
  journal = {A\&A},
  volume = {527},
  pages = {A123},
  doi = {10.1051/0004-6361/201015973}
}

@article{Gusdorf_Probing_2012,
  title = {Probing Magnetohydrodynamic Shocks with High-{{J CO}} Observations: {{W28F}}},
  shorttitle = {Probing Magnetohydrodynamic Shocks with High- {{{\emph{J}}}} {{CO}} Observations},
  author = {Gusdorf, A. and Anderl, S. and G{\"u}sten, R. and Stutzki, J. and H{\"u}bers, H.-W. and Hartogh, P. and Heyminck, S. and Okada, Y.},
  year = 2012,
  journal = {A\&A},
  volume = {542},
  pages = {L19},
  doi = {10.1051/0004-6361/201218907},
  langid = {english}
}

@article{Gusdorf_SiO_2008,
  title = {{{SiO}} Line Emission from {{C-type}} Shock Waves: Interstellar Jets and Outflows},
  shorttitle = {{{SiO}} Line Emission from {{C-type}} Shock Waves},
  author = {Gusdorf, A. and Cabrit, S. and Flower, D. R. and Pineau Des For{\^e}ts, G.},
  year = 2008,
  journal = {A\&A},
  volume = {482},
  number = {3},
  pages = {809--829},
  doi = {10.1051/0004-6361:20078900},
  langid = {english}
}

@article{Harada_Starburst_2021,
  title = {Starburst {{Energy Feedback Seen}} through {{HCO}}+/{{HOC}}+ {{Emission}} in {{NGC}} 253 from {{ALCHEMI}}},
  author = {Harada, Nanase and Mart{\'i}n, Sergio and Mangum, Jeffrey G. and Sakamoto, Kazushi and Muller, Sebastien and Tanaka, Kunihiko and Nakanishi, Kouichiro and {Herrero-Illana}, Rub{\'e}n and Yoshimura, Yuki and M{\"u}hle, Stefanie and Aladro, Rebeca and Colzi, Laura and Rivilla, V{\'i}ctor M. and Aalto, Susanne and Behrens, Erica and Henkel, Christian and Holdship, Jonathan and Humire, P. K. and Meier, David S. and Nishimura, Yuri and {van der Werf}, Paul P. and Viti, Serena},
  year = 2021,
  journal = {ApJ},
  volume = {923},
  pages = {24},
  doi = {10.3847/1538-4357/ac26b8}
}

@article{Hernandez-Hernandez_APEX_2019,
  title = {{{APEX Millimeter Observations}} of {{Methanol Emission Toward High-mass Star-forming Cores}}},
  author = {{Hern{\'a}ndez-Hern{\'a}ndez}, Vicente and Kurtz, Stan and Kalenskii, Sergei and Golysheva, Polina and Garay, Guido and Zapata, Luis and Bergman, Per},
  year = 2019,
  journal = {ApJ},
  volume = {158},
  pages = {18},
  publisher = {IOP},
  doi = {10.3847/1538-3881/ab2047}
}

@article{Hiramatsu_Outflow-Core_2010a,
  title = {Outflow-{{Core Interaction}} in {{Barnard}} 1},
  author = {Hiramatsu, Masaaki and Hirano, Naomi and Takakuwa, Shigehisa},
  year = 2010,
  journal = {The Astrophysical Journal},
  volume = {712},
  pages = {778--789},
  publisher = {IOP},
  doi = {10.1088/0004-637X/712/2/778}
}

@article{Hoffman_OH_2005,
  title = {The {{OH}} (1720 {{MHz}}) {{Supernova Remnant Masers}} in {{W28}}: {{MERLIN}} and {{VLBA Polarization Observations}}},
  shorttitle = {The {{OH}} (1720 {{MHz}}) {{Supernova Remnant Masers}} in {{W28}}},
  author = {Hoffman, Ian M. and Goss, W. M. and Brogan, C. L. and Claussen, M. J.},
  year = 2005,
  journal = {ApJ},
  volume = {620},
  pages = {257--273},
  doi = {10.1086/427018}
}

@article{Hogge_Interaction_2019,
  title = {The {{Interaction}} between the {{Supernova Remnant W41}} and the {{Filamentary Infrared Dark Cloud G23}}.33-0.30},
  author = {Hogge, Taylor G. and Jackson, James M. and Allingham, David and Guzman, Andres E. and {Killerby-Smith}, Nicholas and Kraemer, Kathleen E. and Sanhueza, Patricio and Stephens, Ian W. and Whitaker, J. Scott},
  year = 2019,
  journal = {ApJ},
  volume = {887},
  pages = {79},
  doi = {10.3847/1538-4357/ab5180}
}

@article{Holdship_Energizing_2022,
  title = {Energizing {{Star Formation}}: {{The Cosmic-Ray Ionization Rate}} in {{NGC}} 253 {{Derived}} from {{ALCHEMI Measurements}} of {{H3O}}+ and {{SO}}},
  shorttitle = {Energizing {{Star Formation}}},
  author = {Holdship, Jonathan and Mangum, Jeffrey G. and Viti, Serena and Behrens, Erica and Harada, Nanase and Mart{\'i}n, Sergio and Sakamoto, Kazushi and Muller, Sebastien and Tanaka, Kunihiko and Nakanishi, Kouichiro and {Herrero-Illana}, Rub{\'e}n and Yoshimura, Yuki and Aladro, Rebeca and Colzi, Laura and Emig, Kimberly L. and Henkel, Christian and Nishimura, Yuri and Rivilla, V{\'i}ctor M. and {van der Werf}, Paul P. and {Alma Comprehensive High-Resolution Extragalactic Molecular Inventory (Alchemi) Collaboration}},
  year = 2022,
  journal = {ApJ},
  volume = {931},
  pages = {89},
  doi = {10.3847/1538-4357/ac6753}
}

@article{Holdship_Observations_2019,
  title = {Observations of {{CH3OH}} and {{CH3CHO}} in a {{Sample}} of {{Protostellar Outflow Sources}}},
  author = {Holdship, Jonathan and Viti, Serena and Codella, Claudio and Rawlings, Jonathan and {Jimenez-Serra}, Izaskun and Ayalew, Yenabeb and Curtis, Justin and Habib, Annur and Lawrence, Jamel and Warsame, Sumaya and Horn, Sarah},
  year = 2019,
  journal = {ApJ},
  volume = {880},
  pages = {138},
  publisher = {IOP},
  doi = {10.3847/1538-4357/ab1f8f}
}

@article{Hsieh_PRODIGE_2024a,
  title = {{{PRODIGE}} - Envelope to Disk with {{NOEMA}}. {{III}}. {{The}} Origin of Complex Organic Molecule Emission in {{SVS13A}}},
  author = {Hsieh, T. -H. and Pineda, J. E. and {Segura-Cox}, D. M. and Caselli, P. and {Valdivia-Mena}, M. T. and Gieser, C. and Maureira, M. J. and {Lopez-Sepulcre}, A. and Bouscasse, L. and Neri, R. and M{\"o}ller, {\relax Th}. and Dutrey, A. and Fuente, A. and Semenov, D. and Chapillon, E. and Cunningham, N. and Henning, {\relax Th}. and Pi{\'e}tu, V. and {Jimenez-Serra}, I. and Marino, S. and Ceccarelli, C.},
  year = 2024,
  journal = {A\&A},
  volume = {686},
  pages = {A289},
  publisher = {EDP},
  doi = {10.1051/0004-6361/202449417}
}

@article{Huang_Investigating_2025,
  title = {Investigating the Chemical Link between {{H2CO}} and {{CH3OH}} within the Central Molecular Zone of {{NGC}} 253},
  author = {Huang, K. -Y. and Behrens, E. and Bouvier, M. and Viti, S. and Mangum, J. G. and Eibensteiner, C.},
  year = 2025,
  journal = {A\&A},
  volume = {699},
  pages = {A70},
  publisher = {EDP},
  doi = {10.1051/0004-6361/202554156}
}

@article{Huang_Investigating_2025b,
  title = {Investigating Solid-State {{CH3OH}} Formation with Chemical Modelling},
  author = {Huang, K.-Y. and {M{\'e}ndez-Robayo}, E. and Viti, S. and {Higuera-G.}, M.-A.},
  year = 2025,
  journal = {A\&A},
  volume = {704},
  pages = {A17},
  publisher = {EDP},
  doi = {10.1051/0004-6361/202556145}
}

@article{Huang_Tracing_2023,
  title = {Tracing the Chemical Footprint of Shocks in {{AGN-host}} and Starburst Galaxies with {{ALMA}} Multi-Line Molecular Studies},
  author = {Huang, Ko-Yun and Viti, Serena},
  year = 2023,
  journal = {FaDi},
  volume = {245},
  pages = {181--198},
  doi = {10.1039/D3FD00007A}
}

@article{Indriolo_Absorption-line_2023,
  title = {Absorption-Line {{Observations}} of \textbraceleft\textbraceleft{{H}}\textbraceright\textbraceright\_\textbraceleft 3\textbraceright (+) and {{CO}} in {{Sight Lines Toward}} the {{Vela}} and {{W28 Supernova Remnants}}},
  author = {Indriolo, Nick},
  year = 2023,
  journal = {ApJ},
  volume = {950},
  pages = {64},
  doi = {10.3847/1538-4357/acc6c4}
}

@article{Indriolo_Investigating_2010,
  title = {Investigating the {{Cosmic-ray Ionization Rate Near}} the {{Supernova Remnant IC}} 443 through {{H}}{\textsubscript{3}}{\textsuperscript{+}} {{Observations}}},
  author = {Indriolo, Nick and Blake, Geoffrey A. and Goto, Miwa and Usuda, Tomonori and Oka, Takeshi and Geballe, T. R. and Fields, Brian D. and McCall, Benjamin J.},
  year = 2010,
  journal = {ApJ},
  volume = {724},
  pages = {1357--1365},
  doi = {10.1088/0004-637X/724/2/1357}
}

@article{Jiang_Cavity_2010,
  title = {Cavity of {{Molecular Gas Associated}} with {{Supernova Remnant 3C}} 397},
  author = {Jiang, Bing and Chen, Yang and Wang, Junzhi and Su, Yang and Zhou, Xin and {Safi-Harb}, Samar and DeLaney, Tracey},
  year = 2010,
  journal = {ApJ},
  volume = {712},
  number = {2},
  eprint = {1001.2204},
  pages = {1147--1156},
  doi = {10.1088/0004-637X/712/2/1147},
  archiveprefix = {arXiv},
  langid = {english}
}

@article{Jimenez-Serra_Parametrization_2008,
  title = {Parametrization of {{C-shocks}}. {{Evolution}} of the Sputtering of Grains},
  author = {{Jim{\'e}nez-Serra}, I. and Caselli, P. and {Mart{\'i}n-Pintado}, J. and Hartquist, T. W.},
  year = 2008,
  journal = {A\&A},
  volume = {482},
  number = {2},
  pages = {549--559},
  doi = {10.1051/0004-6361:20078054},
  langid = {english}
}

@article{Jorgensen_impact_2004,
  title = {The Impact of Shocks on the Chemistry of Molecular Clouds. {{High}} Resolution Images of Chemical Differentiation along the {{NGC}} 1333-{{IRAS 2A}} Outflow},
  author = {J{\o}rgensen, J. K. and Hogerheijde, M. R. and Blake, G. A. and {van Dishoeck}, E. F. and Mundy, L. G. and Sch{\"o}ier, F. L.},
  year = 2004,
  journal = {A\&A},
  volume = {415},
  pages = {1021--1037},
  publisher = {EDP},
  doi = {10.1051/0004-6361:20034216}
}

@article{Kim_Momentum_2015,
  title = {Momentum {{Injection}} by {{Supernovae}} in the {{Interstellar Medium}}},
  author = {Kim, Chang-Goo and Ostriker, Eve C.},
  year = 2015,
  journal = {ApJ},
  volume = {802},
  pages = {99},
  publisher = {IOP},
  doi = {10.1088/0004-637X/802/2/99}
}

@article{Krumholz_Star_2009,
  title = {The {{Star Formation Law}} in {{Atomic}} and {{Molecular Gas}}},
  author = {Krumholz, Mark R. and McKee, Christopher F. and Tumlinson, Jason},
  year = 2009,
  journal = {ApJ},
  volume = {699},
  pages = {850--856},
  publisher = {IOP},
  doi = {10.1088/0004-637X/699/1/850}
}

@article{Lefloch_CHESS_2012,
  title = {The {{CHESS Survey}} of the {{L1157-B1 Shock Region}}: {{CO Spectral Signatures}} of {{Jet-driven Bow Shocks}}},
  shorttitle = {The {{CHESS Survey}} of the {{L1157-B1 Shock Region}}},
  author = {Lefloch, B. and Cabrit, S. and Busquet, G. and Codella, C. and Ceccarelli, C. and Cernicharo, J. and Pardo, J. R. and Benedettini, M. and Lis, D. C. and Nisini, B.},
  year = 2012,
  journal = {ApJ},
  volume = {757},
  pages = {L25},
  publisher = {IOP},
  doi = {10.1088/2041-8205/757/2/L25}
}

@article{Lehmann_Self-generated_2022,
  title = {Self-Generated Ultraviolet Radiation in Molecular Shock Waves. {{II}}. {{CH}}+ and the Interpretation of Emission from Shock Ensembles},
  author = {Lehmann, A. and Godard, B. and {Pineau des For{\^e}ts}, G. and {Vidal-Garc{\'i}a}, A. and Falgarone, E.},
  year = 2022,
  journal = {A\&A},
  volume = {658},
  pages = {A165},
  doi = {10.1051/0004-6361/202141487}
}

@article{Lin_Massive_2024a,
  title = {Massive Clumps in {{W43-main}}: {{Structure}} Formation in an Extensively Shocked Molecular Cloud},
  shorttitle = {Massive Clumps in {{W43-main}}},
  author = {Lin, Y. and Wyrowski, F. and Liu, H. B. and Gong, Y. and Sipil{\"a}, O. and Izquierdo, A. and Csengeri, T. and Ginsburg, A. and Li, G. X. and Spezzano, S. and Pineda, J. E. and Leurini, S. and Caselli, P. and Menten, K. M.},
  year = 2024,
  journal = {A\&A},
  volume = {685},
  pages = {A101},
  publisher = {EDP},
  doi = {10.1051/0004-6361/202348959}
}

@article{Lique_Rotationally_2007,
  title = {Rotationally Inelastic Collisions of {{SO}}({{X 3$\Sigma$-}}) with {{H2}}: {{Potential}} Energy Surface and Rate Coefficients for Excitation by Para-{{H2}} at Low Temperature},
  shorttitle = {Rotationally Inelastic Collisions of {{SO}}({{X 3$\Sigma$-}}) with {{H2}}},
  author = {Lique, F. and Senent, M.-L. and Spielfiedel, A. and Feautrier, N.},
  year = 2007,
  journal = {JChPh},
  volume = {126},
  pages = {164312--164312},
  publisher = {AIP},
  doi = {10.1063/1.2723733}
}

@article{Martin-Pintado_SiO_1992,
  title = {{{SiO}} Emission as a Tracer of Shocked Gas in Molecular Outflows.},
  author = {{Martin-Pintado}, J. and Bachiller, R. and Fuente, A.},
  year = 1992,
  journal = {A\&A},
  volume = {254},
  pages = {315--326}
}

@article{Maxted_Ammonia_2016,
  title = {Ammonia Excitation Imaging of Shocked Gas towards the {{W28}} Gamma-Ray Source {{HESS J1801-233}}},
  author = {Maxted, Nigel I. and {de Wilt}, Phoebe and Rowell, Gavin P. and Nicholas, Brent P. and Burton, Michael G. and Walsh, Andrew and Fukui, Yasuo and Kawamura, Akiko},
  year = 2016,
  journal = {MNRAS},
  volume = {462},
  number = {1},
  eprint = {1607.03061},
  pages = {532--546},
  doi = {10.1093/mnras/stw1687},
  archiveprefix = {arXiv},
  langid = {english}
}

@article{Mazumdar_Submillimeter_2022a,
  title = {Submillimeter Observations of Molecular Gas Interacting with the Supernova Remnant {{W28}}},
  author = {Mazumdar, Parichay and Tram, Le Ngoc and Wyrowski, Friedrich and Menten, Karl M. and Tang, Xindi},
  year = 2022,
  journal = {A\&A},
  volume = {668},
  pages = {A180},
  doi = {10.1051/0004-6361/202037564},
  langid = {english}
}

@article{Meijerink_Irradiated_2006,
  title = {Irradiated {{ISM}}: {{Discriminating}} between {{Cosmic Rays}} and {{X-Rays}}},
  shorttitle = {Irradiated {{ISM}}},
  author = {Meijerink, R. and Spaans, M. and Israel, F. P.},
  year = 2006,
  journal = {ApJ},
  volume = {650},
  pages = {L103-L106},
  doi = {10.1086/508938}
}

@article{Menten_Methanol_1988,
  title = {Methanol in the {{Orion}} Region. {{I}}. {{Millimeter-wave}} Observations.},
  author = {Menten, K. M. and Walmsley, C. M. and Henkel, C. and Wilson, T. L.},
  year = 1988,
  journal = {A\&A},
  volume = {198},
  pages = {253--266}
}

@article{Neufeld_Water_2014,
  title = {The {{Water Abundance}} behind {{Interstellar Shocks}}: {{Results}} from {{Herschel}}/{{PACS}} and {{Spitzer}}/{{IRS Observations}} of {{H2O}}, {{CO}}, and {{H2}}},
  shorttitle = {The {{Water Abundance}} behind {{Interstellar Shocks}}},
  author = {Neufeld, David A. and Gusdorf, Antoine and G{\"u}sten, Rolf and Herczeg, Greg J. and Kristensen, Lars and Melnick, Gary J. and Nisini, Brunella and Ossenkopf, Volker and Tafalla, Mario and {van Dishoeck}, Ewine F.},
  year = 2014,
  journal = {ApJ},
  volume = {781},
  pages = {102},
  doi = {10.1088/0004-637X/781/2/102}
}

@article{Nguyen-Luong_Low-velocity_2013,
  title = {Low-Velocity {{Shocks Traced}} by {{Extended SiO Emission}} along the {{W43 Ridges}}: {{Witnessing}} the {{Formation}} of {{Young Massive Clusters}}},
  shorttitle = {Low-Velocity {{Shocks Traced}} by {{Extended SiO Emission}} along the {{W43 Ridges}}},
  author = {{Nguyen-Lu'o'ng}, Q. and Motte, F. and Carlhoff, P. and Louvet, F. and Lesaffre, P. and Schilke, P. and Hill, T. and Hennemann, M. and Gusdorf, A. and Didelon, P. and Schneider, N. and Bontemps, S. and {Duarte-Cabral}, A. and Menten, K. M. and Martin, P. G. and Wyrowski, F. and Bendo, G. and Roussel, H. and Bernard, J. -P. and Bronfman, L. and Henning, T. and Kramer, C. and Heitsch, F.},
  year = 2013,
  journal = {ApJ},
  volume = {775},
  pages = {88},
  publisher = {IOP},
  doi = {10.1088/0004-637X/775/2/88}
}

@article{Nicholas_12_2011,
  title = {12 Mm Line Survey of the Dense Molecular Gas towards the {{W28}} Field {{TeV}} Gamma-Ray Sources: 12 Mm Line Survey: {{W28}} Field {{TeV}} Sources},
  shorttitle = {12 Mm Line Survey of the Dense Molecular Gas towards the {{W28}} Field {{TeV}} Gamma-Ray Sources},
  author = {Nicholas, B. and Rowell, G. and Burton, M. G. and Walsh, A. and Fukui, Y. and Kawamura, A. and Longmore, S. and Keto, E.},
  year = 2011,
  journal = {MNRAS},
  volume = {411},
  number = {2},
  pages = {1367--1385},
  doi = {10.1111/j.1365-2966.2010.17778.x},
  langid = {english}
}

@article{Nicholas_7_2012,
  title = {A 7 Mm Line Survey of the Shocked and Disrupted Molecular Gas towards the {{W28}} Field {{TeV}} Gamma-Ray Sources: 7 Mm Line Survey: {{W28}} Field {{TeV}} Sources},
  shorttitle = {A 7 Mm Line Survey of the Shocked and Disrupted Molecular Gas towards the {{W28}} Field {{TeV}} Gamma-Ray Sources},
  author = {Nicholas, B. P. and Rowell, G. and Burton, M. G. and Walsh, A. J. and Fukui, Y. and Kawamura, A. and Maxted, N. I.},
  year = 2012,
  journal = {MNRAS},
  volume = {419},
  number = {1},
  pages = {251--266},
  doi = {10.1111/j.1365-2966.2011.19688.x},
  langid = {english}
}

@article{Olmi_Herschel-HIFI_2015,
  title = {Herschel-{{HIFI}} Observations of {{H2O}}, {{NH3}}, and {{N2H}}+ toward High-Mass Starless and Protostellar Clumps Identified by the {{Hi-GAL}} Survey{$\star$}},
  author = {Olmi, L. and Persson, C. M. and Codella, C.},
  year = 2015,
  journal = {A\&A},
  volume = {583},
  pages = {A125},
  publisher = {EDP},
  doi = {10.1051/0004-6361/201526901}
}

@article{Pihlstrom_Detection_2014,
  title = {Detection of {{Class I Methanol}} ({{CH3OH}}) {{Maser Candidates}} in {{Supernova Remnants}}},
  author = {Pihlstr{\"o}m, Y. M. and Sjouwerman, L. O. and Frail, D. A. and Claussen, M. J. and Mesler, R. A. and McEwen, B. C.},
  year = 2014,
  journal = {ApJ},
  volume = {147},
  pages = {73},
  doi = {10.1088/0004-6256/147/4/73}
}

@article{Pillepich_Simulating_2018,
  title = {Simulating Galaxy Formation with the {{IllustrisTNG}} Model},
  author = {Pillepich, Annalisa and Springel, Volker and Nelson, Dylan and Genel, Shy and Naiman, Jill and Pakmor, R{\"u}diger and Hernquist, Lars and Torrey, Paul and Vogelsberger, Mark and Weinberger, Rainer and Marinacci, Federico},
  year = 2018,
  journal = {MNRAS},
  volume = {473},
  pages = {4077--4106},
  doi = {10.1093/mnras/stx2656}
}

@article{PineaudesForets_Sulphur-bearing_1993,
  title = {Sulphur-Bearing Molecules as Tracers of Shocks in Interstellar Clouds},
  author = {{Pineau des Forets}, G. and Roueff, E. and Schilke, P. and Flower, D. R.},
  year = 1993,
  journal = {MNRAS},
  volume = {262},
  pages = {915--928},
  publisher = {OUP},
  doi = {10.1093/mnras/262.4.915}
}

@article{Plunkett_Data_2023a,
  title = {Data {{Combination}}: {{Interferometry}} and {{Single-dish Imaging}} in {{Radio Astronomy}}},
  shorttitle = {Data {{Combination}}},
  author = {Plunkett, Adele and Hacar, Alvaro and {Moser-Fischer}, Lydia and Petry, Dirk and Teuben, Peter and Pingel, Nickolas and Kunneriath, Devaky and Takagi, Toshinobu and Miyamoto, Yusuke and Moravec, Emily and Suri, S{\"u}meyye and Hess, Kelley M. and Hoffman, Melissa and Mason, Brian},
  year = 2023,
  journal = {PASA},
  volume = {135},
  pages = {034501},
  publisher = {IOP},
  doi = {10.1088/1538-3873/acb9bd}
}

@article{Purcell_Physical_2009,
  title = {Physical and Chemical Conditions in Methanol Maser Selected Hot Cores and {{UCHII}} Regions},
  author = {Purcell, C. R. and Longmore, S. N. and Burton, M. G. and Walsh, A. J. and Minier, V. and Cunningham, M. R. and Balasubramanyam, R.},
  year = 2009,
  journal = {Monthly Notices of the Royal Astronomical Society},
  volume = {394},
  pages = {323--339},
  publisher = {OUP},
  doi = {10.1111/j.1365-2966.2008.14283.x}
}

@article{Reach_Excitation_1999,
  title = {Excitation and {{Disruption}} of a {{Giant Molecular Cloud}} by the {{Supernova Remnant 3C}} 391},
  author = {Reach, William T. and Rho, Jeonghee},
  year = 1999,
  journal = {ApJ},
  volume = {511},
  pages = {836--846},
  doi = {10.1086/306703}
}

@article{Reach_Shocked_2005,
  title = {Shocked {{Molecular Gas}} in the {{Supernova Remnants W28}} and {{W44}}: {{Near}}-{{Infrared}} and {{Millimeter}}-{{Wave Observations}}},
  shorttitle = {Shocked {{Molecular Gas}} in the {{Supernova Remnants W28}} and {{W44}}},
  author = {Reach, William T. and Rho, Jeonghee and Jarrett, T. H.},
  year = 2005,
  journal = {ApJ},
  volume = {618},
  number = {1},
  pages = {297--320},
  doi = {10.1086/425855},
  langid = {english}
}

@article{Reach_Supernova_2019,
  title = {Supernova {{Shocks}} in {{Molecular Clouds}}: {{Velocity Distribution}} of {{Molecular Hydrogen}}},
  shorttitle = {Supernova {{Shocks}} in {{Molecular Clouds}}},
  author = {Reach, William T. and Tram, Le Ngoc and Richter, Matthew and Gusdorf, Antoine and DeWitt, Curtis},
  year = 2019,
  journal = {ApJ},
  volume = {884},
  pages = {81},
  doi = {10.3847/1538-4357/ab41f7}
}

@article{Reach_Supernova_2024,
  title = {Supernova {{Shocks}} in {{Molecular Clouds}}: {{Shocks Driven}} into {{Dense Cores}} in {{IC}} 443 and {{3C}} 391},
  shorttitle = {Supernova {{Shocks}} in {{Molecular Clouds}}},
  author = {Reach, William T. and Tram, Le Ngoc and DeWitt, Curtis and Lesaffre, Pierre and Godard, Benjamin and Gusdorf, Antoine},
  year = 2024,
  journal = {ApJ},
  volume = {977},
  pages = {149},
  publisher = {IOP},
  doi = {10.3847/1538-4357/ad8d59}
}

@article{Rho_Detection_2015,
  title = {Detection of {{Extremely Broad Water Emission}} from the {{Molecular Cloud Interacting Supernova Remnant G349}}.7+0.2},
  author = {Rho, J. and Hewitt, J. W. and Boogert, A. and Kaufman, M. and Gusdorf, A.},
  year = 2015,
  journal = {ApJ},
  volume = {812},
  pages = {44},
  doi = {10.1088/0004-637X/812/1/44}
}

@article{Rho_ROSAT_2002,
  title = {{{ROSAT}}/{{ASCA Observations}} of the {{Mixed-Morphology Supernova Remnant W28}}},
  author = {Rho, Jeonghee and Borkowski, Kazimierz J.},
  year = 2002,
  journal = {ApJ},
  volume = {575},
  pages = {201--216},
  doi = {10.1086/341192}
}

@article{Rodriguez-Fernandez_HNCO_2010,
  title = {{{HNCO}} Enhancement by Shocks in the {{L1157}} Molecular Outflow},
  author = {{Rodr{\'i}guez-Fern{\'a}ndez}, N. J. and Tafalla, M. and Gueth, F. and Bachiller, R.},
  year = 2010,
  journal = {A\&A},
  volume = {516},
  pages = {A98},
  publisher = {EDP},
  doi = {10.1051/0004-6361/201013997}
}

@article{Sabatini_FAUST_2024,
  title = {{{FAUST}}. {{XIII}}. {{Dusty}} Cavity and Molecular Shock Driven by {{IRS7B}} in the {{Corona Australis}} Cluster},
  author = {Sabatini, G. and Podio, L. and Codella, C. and Watanabe, Y. and De Simone, M. and Bianchi, E. and Ceccarelli, C. and Chandler, C. J. and Sakai, N. and Svoboda, B. and Testi, L. and Aikawa, Y. and Balucani, N. and Bouvier, M. and Caselli, P. and Caux, E. and Chahine, L. and Charnley, S. and Cuello, N. and Dulieu, F. and Evans, L. and Fedele, D. and Feng, S. and Fontani, F. and Hama, T. and Hanawa, T. and Herbst, E. and Hirota, T. and Isella, A. and {J{\'i}menez-Serra}, I. and Johnstone, D. and Lefloch, B. and Le Gal, R. and Loinard, L. and Liu, H. B. and {L{\'o}pez-Sepulcre}, A. and Maud, L. T. and Maureira, M. J. and Menard, F. and Miotello, A. and Moellenbrock, G. and Nomura, H. and Oba, Y. and Ohashi, S. and Okoda, Y. and Oya, Y. and Pineda, J. and Rimola, A. and Sakai, T. and {Segura-Cox}, D. and Shirley, Y. and Vastel, C. and Viti, S. and Watanabe, N. and Zhang, Y. and Zhang, Z. E. and Yamamoto, S.},
  year = 2024,
  journal = {A\&A},
  volume = {684},
  pages = {L12},
  publisher = {EDP},
  doi = {10.1051/0004-6361/202449616}
}

@article{Schilke_SiO_1997,
  title = {{{SiO}} Production in Interstellar Shocks.},
  author = {Schilke, P. and Walmsley, C. M. and {Pineau des Forets}, G. and Flower, D. R.},
  year = 1997,
  journal = {A\&A},
  volume = {321},
  pages = {293--304}
}

@article{Shirley_Critical_2015,
  title = {The {{Critical Density}} and the {{Effective Excitation Density}} of {{Commonly Observed Molecular Dense Gas Tracers}}},
  author = {Shirley, Yancy L.},
  year = 2015,
  journal = {PASP},
  volume = {127},
  pages = {299},
  doi = {10.1086/680342}
}

@article{Simons_Formation_2020,
  title = {Formation of {{COMs}} through {{CO}} Hydrogenation on Interstellar Grains},
  author = {Simons, M. A. J. and Lamberts, T. and Cuppen, H. M.},
  year = 2020,
  journal = {A\&A},
  volume = {634},
  pages = {A52},
  publisher = {EDP},
  doi = {10.1051/0004-6361/201936522}
}

@article{Suutarinen_Water_2014,
  title = {Water and Methanol in Low-Mass Protostellar Outflows: Gas-Phase Synthesis, Ice Sputtering and Destruction},
  shorttitle = {Water and Methanol in Low-Mass Protostellar Outflows},
  author = {Suutarinen, A. N. and Kristensen, L. E. and Mottram, J. C. and Fraser, H. J. and {van Dishoeck}, E. F.},
  year = 2014,
  journal = {MNRAS},
  volume = {440},
  pages = {1844--1855},
  publisher = {OUP},
  doi = {10.1093/mnras/stu406}
}

@article{Tafalla_HH_2013,
  title = {{{HH}} 114 {{MMS}}: A New Chemically Active Outflow},
  shorttitle = {{{HH}} 114 {{MMS}}},
  author = {Tafalla, M. and Hacar, A.},
  year = 2013,
  journal = {A\&A},
  volume = {552},
  pages = {L9},
  publisher = {EDP},
  doi = {10.1051/0004-6361/201321303}
}

@article{Tafalla_molecular_2010,
  title = {A Molecular Survey of Outflow Gas: Velocity-Dependent Shock Chemistry and the Peculiar Composition of the {{EHV}} Gas},
  shorttitle = {A Molecular Survey of Outflow Gas},
  author = {Tafalla, M. and {Santiago-Garc{\'i}a}, J. and Hacar, A. and Bachiller, R.},
  year = 2010,
  journal = {A\&A},
  volume = {522},
  pages = {A91},
  doi = {10.1051/0004-6361/201015158}
}

@article{Tu_Shock_2024,
  title = {Shock and {{Cosmic-Ray Chemistry Associated}} with the {{Supernova Remnant W28}}},
  author = {Tu, Tian-yu and Chen, Yang and Zhou, Ping and {Safi-Harb}, Samar and Liu, Qian-Cheng},
  year = 2024,
  journal = {ApJ},
  volume = {966},
  pages = {178},
  publisher = {IOP},
  doi = {10.3847/1538-4357/ad3634}
}

@article{Tu_Yebes_2024,
  title = {A {{Yebes W-band Line Survey}} towards an {{Unshocked Molecular Cloud}} of {{Supernova Remnant 3C}} 391: {{Evidence}} of {{Cosmic-Ray-Induced Chemistry}}},
  shorttitle = {A {{Yebes W-band Line Survey}} towards an {{Unshocked Molecular Cloud}} of {{Supernova Remnant 3C}} 391},
  author = {Tu, Tian-Yu and Rayalacheruvu, Prathap and Majumdar, Liton and Chen, Yang and Zhou, Ping and {Santander-Garc{\'i}a}, Miguel},
  year = 2024,
  journal = {ApJ},
  volume = {974},
  pages = {262},
  publisher = {IOP},
  doi = {10.3847/1538-4357/ad74fb}
}

@article{Tychoniec_Which_2021,
  title = {Which Molecule Traces What: {{Chemical}} Diagnostics of Protostellar Sources},
  shorttitle = {Which Molecule Traces What},
  author = {Tychoniec, {\L}ukasz and {van Dishoeck}, Ewine F. and {van't Hoff}, Merel L. R. and {van Gelder}, Martijn L. and Tabone, Beno{\^i}t and Chen, Yuan and Harsono, Daniel and Hull, Charles L. H. and Hogerheijde, Michiel R. and Murillo, Nadia M. and Tobin, John J.},
  year = 2021,
  journal = {A\&A},
  volume = {655},
  pages = {A65},
  doi = {10.1051/0004-6361/202140692},
  langid = {english}
}

@article{vanderTak_computer_2007,
  title = {A Computer Program for Fast Non-{{LTE}} Analysis of Interstellar Line Spectra: {{With}} Diagnostic Plots to Interpret Observed Line Intensity Ratios},
  shorttitle = {A Computer Program for Fast Non-{{LTE}} Analysis of Interstellar Line Spectra},
  author = {{van der Tak}, F. F. S. and Black, J. H. and Sch{\"o}ier, F. L. and Jansen, D. J. and {van Dishoeck}, E. F.},
  year = 2007,
  journal = {A\&A},
  volume = {468},
  number = {2},
  pages = {627--635},
  doi = {10.1051/0004-6361:20066820},
  langid = {english}
}

@article{vanderTak_Leiden_2020,
  title = {The {{Leiden Atomic}} and {{Molecular Database}} ({{LAMDA}}): {{Current Status}}, {{Recent Updates}}, and {{Future Plans}}},
  shorttitle = {The {{Leiden Atomic}} and {{Molecular Database}} ({{LAMDA}})},
  author = {{van der Tak}, Floris F. S. and Lique, Fran{\c c}ois and Faure, Alexandre and Black, John H. and {van Dishoeck}, Ewine F.},
  year = 2020,
  journal = {Atoms},
  volume = {8},
  pages = {15},
  doi = {10.3390/atoms8020015}
}

@article{vanDishoeck_Submillimeter_1993,
  title = {Submillimeter Obseravtions of the Shocked Molecular Gas Associated with the Supernova Remnant {{IC}} 443.},
  author = {{van Dishoeck}, E. F. and Jansen, D. J. and Phillips, T. G.},
  year = 1993,
  journal = {A\&A},
  volume = {279},
  pages = {541--566},
  langid = {english}
}

@article{Vaupre_Cosmic_2014,
  title = {Cosmic Ray Induced Ionisation of a Molecular Cloud Shocked by the {{W28}} Supernova Remnant},
  author = {Vaupr{\'e}, S. and {Hily-Blant}, P. and Ceccarelli, C. and Dubus, G. and Gabici, S. and Montmerle, T.},
  year = 2014,
  journal = {A\&A},
  volume = {568},
  pages = {A50},
  doi = {10.1051/0004-6361/201424036},
  langid = {english}
}

@article{Velazquez_Investigation_2002,
  title = {Investigation of the {{Large-scale Neutral Hydrogen}} near the {{Supernova Remnant W28}}},
  author = {Vel{\'a}zquez, P. F. and Dubner, G. M. and Goss, W. M. and Green, A. J.},
  year = 2002,
  journal = {AJ},
  volume = {124},
  pages = {2145--2151},
  doi = {10.1086/342936}
}

@article{Wakelam_Efficiency_2021,
  title = {Efficiency of Non-Thermal Desorptions in Cold-Core Conditions. {{Testing}} the Sputtering of Grain Mantles Induced by Cosmic Rays},
  author = {Wakelam, V. and Dartois, E. and Chabot, M. and Spezzano, S. and {Navarro-Almaida}, D. and Loison, J.-C. and Fuente, A.},
  year = 2021,
  journal = {A\&A},
  volume = {652},
  pages = {A63},
  doi = {10.1051/0004-6361/202039855},
  langid = {english}
}

@article{Watanabe_Hydrogenation_2004,
  title = {Hydrogenation of {{CO}} on {{Pure Solid CO}} and {{CO-H2O Mixed Ice}}},
  author = {Watanabe, Naoki and Nagaoka, Akihiro and Shiraki, Takahiro and Kouchi, Akira},
  year = 2004,
  journal = {ApJ},
  volume = {616},
  pages = {638--642},
  publisher = {IOP},
  doi = {10.1086/424815}
}

@article{Wiesenfeld_Rotational_2013,
  title = {Rotational Quenching of {{H2CO}} by Molecular Hydrogen: Cross-Sections, Rates and Pressure Broadening},
  shorttitle = {Rotational Quenching of {{H2CO}} by Molecular Hydrogen},
  author = {Wiesenfeld, L. and Faure, A.},
  year = 2013,
  journal = {MNRAS},
  volume = {432},
  pages = {2573--2578},
  publisher = {OUP},
  doi = {10.1093/mnras/stt616}
}

@article{Wirstrom_Observational_2011,
  title = {Observational Tests of Interstellar Methanol Formation},
  author = {Wirstr{\"o}m, E. S. and Geppert, W. D. and Hjalmarson, {\AA}. and Persson, C. M. and Black, J. H. and Bergman, P. and Millar, T. J. and Hamberg, M. and Vigren, E.},
  year = 2011,
  journal = {A\&A},
  volume = {533},
  pages = {A24},
  doi = {10.1051/0004-6361/201116525}
}

@article{Wootten_dense_1981,
  title = {A Dense Molecular Cloud Impacted by the {{W28}} Supernova Remnant},
  author = {Wootten, A.},
  year = 1981,
  journal = {ApJ},
  volume = {245},
  pages = {105--114},
  doi = {10.1086/158790}
}

@article{Yamagishi_Cosmic-ray-driven_2023,
  title = {Cosmic-Ray-Driven Enhancement of the {{C0}}/{{CO}} Abundance Ratio in {{W}} 51 {{C}}},
  author = {Yamagishi, Mitsuyoshi and Furuya, Kenji and Sano, Hidetoshi and Izumi, Natsuko and Takekoshi, Tatsuya and Kaneda, Hidehiro and Nakanishi, Kouichiro and Shimonishi, Takashi},
  year = 2023,
  journal = {PASJ},
  volume = {75},
  pages = {883--892},
  doi = {10.1093/pasj/psad046}
}

@article{Yuan_Spitzer_2011,
  title = {Spitzer {{Observations}} of {{Supernova Remnants}}. {{II}}. {{Physical Conditions}} and {{Comparison}} with {{HH7}} and {{HH54}}},
  author = {Yuan, Yuan and Neufeld, David A.},
  year = 2011,
  journal = {ApJ},
  volume = {726},
  pages = {76},
  doi = {10.1088/0004-637X/726/2/76}
}

@article{Zhang_CO_2010,
  title = {{{CO}} Observation of {{SNR IC}} 443},
  author = {Zhang, ZhiYu and Gao, Yu and Wang, JunZhi},
  year = 2010,
  journal = {SCPMA},
  volume = {53},
  pages = {1357--1369},
  publisher = {Springer},
  doi = {10.1007/s11433-010-4010-5}
}

@article{Zhou_Systematic_2023,
  title = {A {{Systematic Study}} of {{Associations}} between {{Supernova Remnants}} and {{Molecular Clouds}}},
  author = {Zhou, Xin and Su, Yang and Yang, Ji and Chen, Xuepeng and Sun, Yan and Jiang, Zhibo and Wang, Min and Wang, Hongchi and Zhang, Shaobo and Xu, Ye and Yan, Qingzeng and Yuan, Lixia and Chen, Zhiwei and Ao, Yiping and Ma, Yuehui},
  year = 2023,
  journal = {ApJS},
  volume = {268},
  pages = {61},
  doi = {10.3847/1538-4365/acee7f}
}

@article{Zhou_Unusually_2022b,
  title = {Unusually {{High HCO}}+/{{CO Ratios}} in and Outside {{Supernova Remnant W49B}}},
  author = {Zhou, Ping and Zhang, Gao-Yuan and Zhou, Xin and Arias, Maria and Koo, Bon-Chul and Vink, Jacco and Zhang, Zhi-Yu and Sun, Lei and Du, Fu-Jun and Zhu, Hui and Chen, Yang and Bovino, Stefano and Lee, Yong-Hyun},
  year = 2022,
  journal = {ApJ},
  volume = {931},
  pages = {144},
  doi = {10.3847/1538-4357/ac63b5}
}

@article{Zhou_XMM-Newton_2014,
  title = {An {{XMM-Newton Study}} of the {{Mixed-morphology Supernova Remnant W28}} ({{G6}}.4-0.1)},
  author = {Zhou, Ping and {Safi-Harb}, Samar and Chen, Yang and Zhang, Xiao and Jiang, Bing and Ferrand, Gilles},
  year = 2014,
  journal = {ApJ},
  volume = {791},
  pages = {87},
  doi = {10.1088/0004-637X/791/2/87}
}
\bibliographystyle{aasjournal}

\end{CJK*}
\end{document}